\documentclass[10pt,journal,cspaper,compsoc]{IEEEtran}
\ifCLASSOPTIONcompsoc
\else
\fi
\ifCLASSINFOpdf
\else
\fi
\usepackage[cmex10]{amsmath}
\usepackage{amssymb}
\providecommand{\shortcite}[1]{\cite{#1}}
\usepackage[pdftex]{graphicx} 
\usepackage{textcomp}
\usepackage{algorithm2e}
\usepackage{subfig}
\usepackage{authblk}

\makeatletter
\newcommand{\clearsubcaptcounter}{\setcounter{sub\@captype}{0}}
\makeatother

\begin{document}
%
\title{Tracing Analytic Ray Curves for Light and Sound Propagation in Non-linear Media}
%
%
%
%

\author[*]{Qi Mo}
\author[*]{Hengchin Yeh}
\author[*]{Dinesh Manocha}
\affil[*]{Department of Computer Science, University of North Carolina, Chapel Hill}
\IEEEcompsoctitleabstractindextext{%
\begin{abstract}
The physical world consists of spatially varying media, such as the atmosphere and the ocean, in which light and sound propagates along non-linear trajectories. This presents a challenge to existing ray-tracing based methods, which are widely adopted to simulate propagation due to their efficiency and flexibility, but assume linear rays. We present a novel algorithm that traces analytic ray curves computed from local media gradients, and utilizes the closed-form solutions of both the intersections of the ray curves with planar surfaces, and the travel distance. By constructing an adaptive unstructured mesh, our algorithm is able to model general media profiles that vary in three dimensions with complex boundaries consisting of terrains and other scene objects such as buildings. We trace the analytic ray curves using the adaptive unstructured mesh, which considerably improves the efficiency over prior methods. We highlight the algorithm's application on simulation of sound and visual propagation in outdoor scenes.

\end{abstract}

}

\maketitle

\IEEEdisplaynotcompsoctitleabstractindextext

%
\IEEEpeerreviewmaketitle

\section{Introduction}
Non-linear media is ubiquitous in the physical world. The atmosphere, even under stable conditions, has spatially varying temperature, pressure, and humidity \cite{usgpc1976standard}. There can be wind field or other weather patterns that affect the atmosphere \cite{monin1957basic,businger1971flux,panofsky1984atmospheric}. Similarly, the ocean displays spatial variations in its key properties such as temperature, pressure, and salinity \cite{jensen2011}. The propagation speed of sound or light wave at a particular location is determined by the spatially varying properties of the media. Refraction refers to the change of propagation direction of a sound or light wave because of a speed gradient; propagation no longer follows linear paths under refraction. Such refractive media is therefore also known as \textit{non-linear media}, and simulating propagation of light and sound in non-linear media remains a challenging problem.

\begin{figure*}
\centering
\subfloat[upward refraction]{\includegraphics[width=0.5\columnwidth]{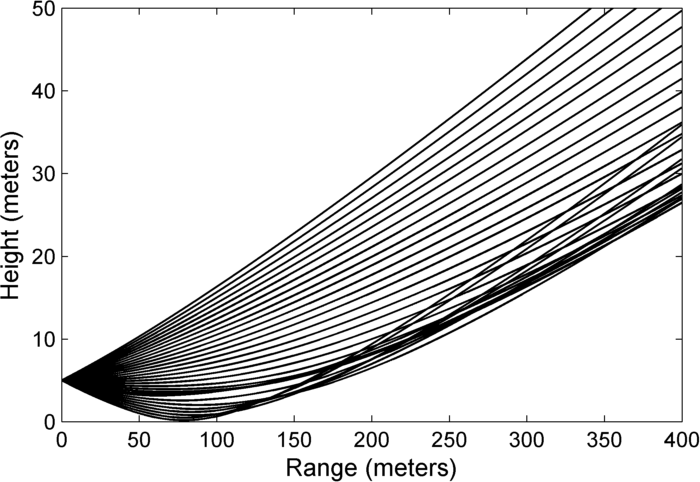}
\label{result:2dray:a}}
\subfloat[hot spot (2D view)]{\includegraphics[width=0.5\columnwidth]{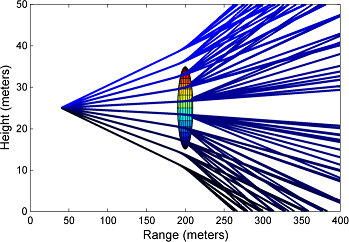}
\label{result:2dray:b}}
\subfloat[upwind over hill (2D view)]{\includegraphics[width=0.5\columnwidth]{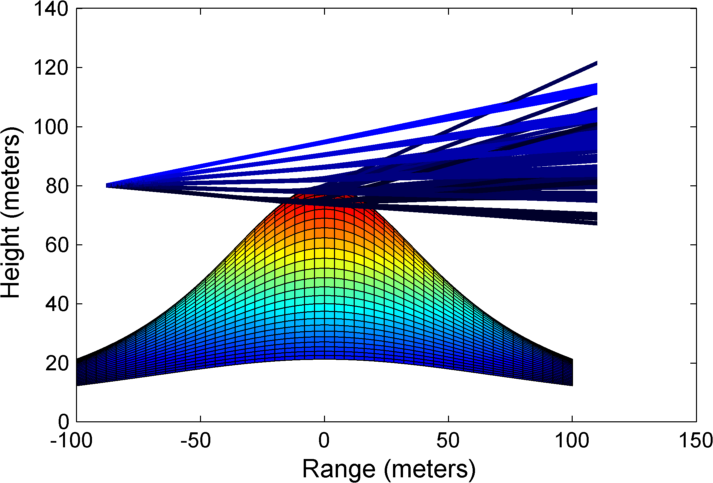}
\label{result:2dray:c}}
\subfloat[downwind over hill (2D view)]{\includegraphics[width=0.5\columnwidth]{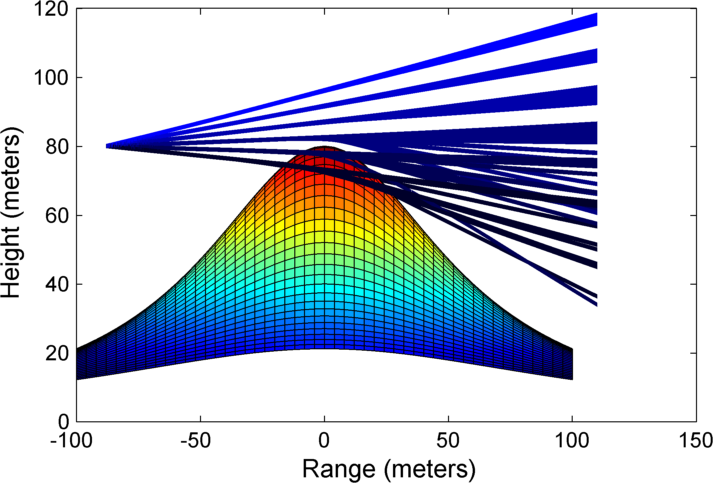}
\label{result:2dray:d}}
\\
\subfloat[downward refraction]{\includegraphics[width=0.5\columnwidth]{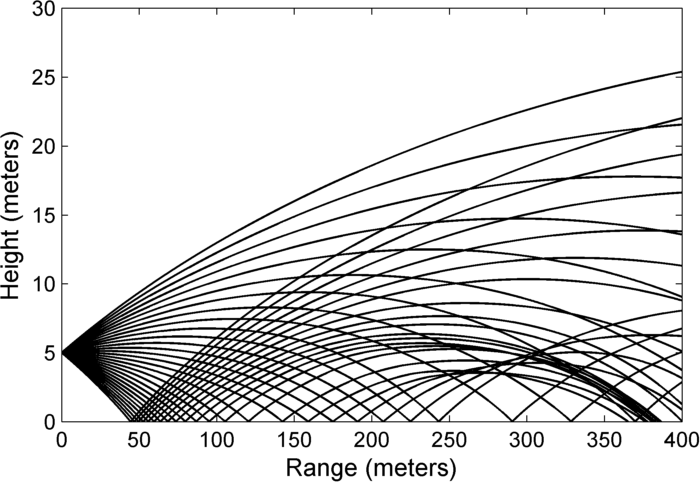}
\label{result:2dray:e}}
\subfloat[hot spot (3D view)]{\includegraphics[width=0.5\columnwidth]{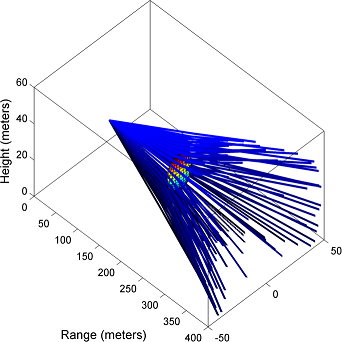}
\label{result:2dray:f}}
\subfloat[upwind over hill (3D view)]{\includegraphics[width=0.5\columnwidth]{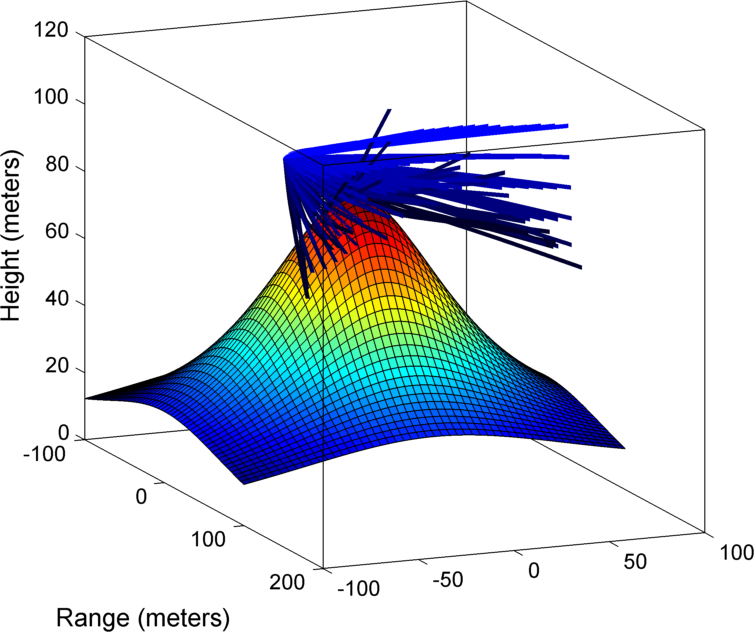}
\label{result:2dray:g}}
\subfloat[downwind over hill (3D view)]{\includegraphics[width=0.5\columnwidth]{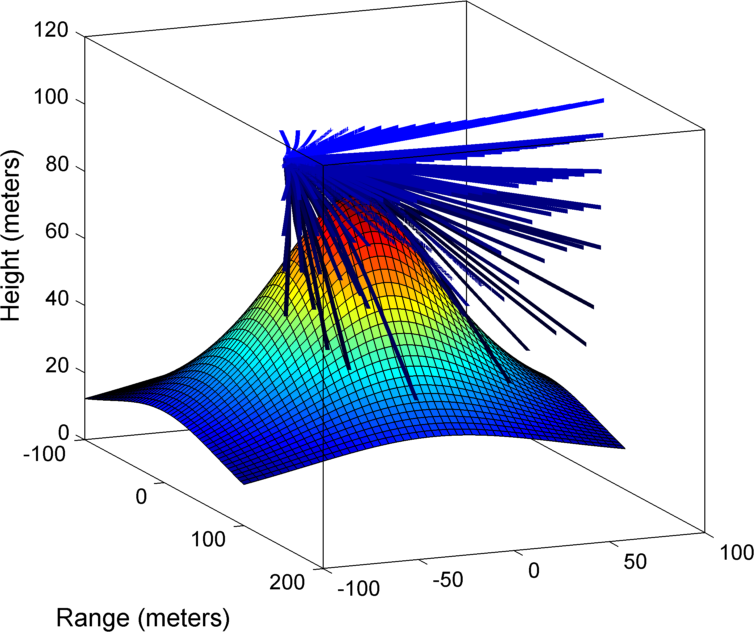}
\label{result:2dray:h}}
\caption{\textbf{Acoustic propagation.} Curved ray trajectories under different atmospheric conditions. The media profiles are generated from physically-based models (see Section \ref{section6-1} for details), including \textbf{(a)} Upward refraction (\textbf{A-LU)} (typical day-time condition), \textbf{(e)} Downward refraction (\textbf{A-LD)} (typical night-time condition), \textbf{(b,f)} Hot spot (\textbf{A-HS}) (the sphere shows the location and influence region of a heat source), \textbf{(c,g)} Up-wind propagation (\textbf{A-UW}) and \textbf{(d,h)} Downwind propagation over a hill (\textbf{A-DW}). The acoustic propagation trajectories deviate significantly from linear paths, and we show the out-of-plane propagation for A-HS, A-UW, and A-DW each from two different views. The 3D varying media profiles lead to complex acoustic fields. Our curved ray tracer computes those paths accurately at 10$\times$ the speed of linear ray stepping.}
\label{result:2dray}
\end{figure*}

Non-linear media in outdoor environments lead to significant acoustic effects \cite{salomons2001}. Take the diurnal change of sound propagation as an example: during the day, when the temperature is typically higher closer to the ground, sound waves are refracted upward, creating a shadow zone with very low level received sound (Figure \ref{result:2dray:a}); when the temperature gradient is inverted at night, sound waves are refracted downward, intensifying the acoustic signals received by the listener. Downward refraction combined with a reflective ground creates a set of concentric circular patterns in the sound field around a source (Figure \ref{result:2dray:e}). Outdoor acoustic applications such as noise reduction, urban planning, and outdoor virtual reality for military training require the propagation simulation to account for those phenomena \cite{attenborough2006predicting,kang2006urban}.

Because the light speed is much larger than sound speed, the non-linear propagation of light in outdoor scenes only becomes apparent under certain conditions (e.g. the extreme temperature gradients that produce mirages \cite{berger1990,musgrave1990,zhao07}.) However, for applications with high accuracy requirements, such as satellite laser range-finding \cite{degnan1993millimeter,dodson1986refraction,gardner1977correction} and solar radiation modeling \cite{badescu2008modeling}, simulating the non-linear propagation paths becomes critical. 

Ray tracing is a powerful tool for simulating sound and light propagation. Traditionally, most ray tracing algorithms focus on linear propagation paths that change directions only at boundary surfaces \cite{Glassner:1989}. Many previous works (See Section \ref{review-piecewiselinear}) adapt the linear ray tracer for non-linear propagation by taking piecewise linear ray steps, effectively assuming a constant media within each linear step. The size of the ray steps therefore becomes seriously limited by the magnitude of variations within the media, hindering the performance of propagating in nonlinear media over long distance. Cao et al. \cite{Cao2010} applied analytic ray formulation from geometric optics to visual rendering based on a locally constant refractive index, which shows promising performance advantage over ray stepping. However, their work does not target large scale general media like the atmosphere, neither has it fully explored the challenges of efficient propagation of both light and sound in complex outdoor scenes.

Some of the prior models and simulator for acoustic propagation  \cite{RTpackages,jensen2011,salomons2001} rely on the assumption of a stratified media, or a media profile that only varies in height and range, reducing the dimension of the problem and making the computation more practical. Given such assumptions, the propagation can even be confined to a 2D plane to reduce the computational overhead, if the media boundary can also be kept simple, i.e. no complex 3D objects to reflect the sound waves off the propagation plane. In reality the media profiles vary in a general manner, and are often altered significantly by complex-shaped 3D objects like buildings or terrains (Figure \ref{result:2dray}(b-d,f-h)). 

\noindent \textbf{Main Results:} In this paper, we present a fast algorithm that traces analytic ray curves for propagation in non-linear media. Compared to the existing methods, our algorithm achieves improvements in the following ways:

\begin{itemize}
\item We trace analytic ray curves as path primitives, which overcomes the step size limitations with linear rays. This is essentially an extension of the idea in \cite{Cao2010}, but we use different ray curve formulations that were derived in the fields of acoustics and optics (see Section \ref{review:analytic} for a full review).
\item We utilize the ray curve formulations (resulting in parabolic or circular rays) to perform closed-form intersections with complex 3D objects, enabling fast propagation in large outdoor scenes with many obstacles.
\item We construct adaptive unstructured tetrahedral mesh based on the underlying media profiles, and we make the media mesh conform to boundaries of scene objects, both of which improve the efficiency of ray curve traversal. 
\end{itemize}

With this algorithm we are able to trace nonlinear sound and light propagation paths for simulation of large and complex outdoor scenes with a general media profiles. We achieve interactive performance on a single CPU core (See Section \ref{section6-3}), and therefore avoid making simplifying assumptions about the media or the scene as made by previous methods for the sake of computational tractability.

\section{Prior Work}

The literature on ray tracing and its acceleration is vast because of its wide range of applications, including photorealistic rendering, geometric acoustics, and scientific visualization. We divide the discussion along two challenges for propagation in non-linear media: (1) computing the curved propagation paths, and (2) characterizing the spatially varying media. In addition, we give an overview of prior work that used closed-form ray curves.

\subsection{Piecewise linear propagation paths}
\label{review-piecewiselinear}

Early works in computer graphics \cite{berger1990,musgrave1990} simulated atmospheric phenomena by modeling the atmosphere with discrete layers. More general media is handled by effectively tracing linear ray segments at each step of a numerical solution of the differential ray equation, derived from either Eikonal equation \cite{Stam96,BHBDGPS12} or Fermat's principle \cite{GSMA06,Satoh03}. Similar methods \cite{Groller1995,Weiskopf2000} have been proposed for modeling gravitational fields and dynamic systems. Piecewise linear approximation of curved paths are also at the heart of techniques such as non-linear photon mapping \cite{GMAS05}, explicit wavefront tracking \cite{Ihrke2007,zhao07}, and voxel-based ray marching \cite{Sun:2008:IRD:1360612.1360634}. Acceleration has been achieved by parallelism \cite{WSE04,zhao07,Ihrke2007}, and spatial and temporal caching \cite{MunozGS07}. 

However, the step size of linear ray tracing is inherently limited by the magnitude of media variations, hindering the scalability of these methods with physical size and complexity of the media and the scenes. Higher order numerical methods like the fourth-order Runge-Kutta are adopted to improve the efficiency \cite{GSMA06,Satoh03,Groller1995,Weiskopf2000,GMAS05}, but the step size is still limited by the underlying media profiles. Furthermore, each advancement of the ray step with higher order numerical methods can no longer be assumed to be a straight line, making intersection tests with the scenes more complex.

In atmospheric and underwater acoustics, seismic modeling, and related fields, similar techniques for tracing piece-wise linear paths have been proposed (see \cite{salomons2001,jensen2011} for a comprehensive survey) and adopted in practical tools \cite{RTpackages}. Just as in computer graphics, the small ray step size becomes a bottleneck; with some of the widely-used software (e.g. BELLHOP) \cite{RTpackages}, simplifying assumptions like a 2D variation of the media (media profiles only vary with height and range), or 2D objects (e.g. conical hills) are often made to keep computation costs feasible.

\subsection{Data structures for non-linear media}
Traditional ray tracing acceleration focuses on building and updating tight-fitting hierarchical structures to enclose only the surfaces in the scenes (see surveys \cite{Glassner:1989,Havran2000}), given a homogeneous media assumption. A noted exception is the use of constrained Delaunay tetrahedralization (CDT) by Lagae and Dutr\'e~\cite{LD08ARTCT}, which adapts to the density of surfaces in the scene without being hierarchical. In contrast, development of efficient techniques for participating media faces the same challenge as that of simulating non-linear media: both must characterize volumetric media in addition to surfaces (see surveys \cite{Cerezo2005,gutierrez09}). Adaptive structures such as kd-trees \cite{Yue:2010,journals/tog/Museth13}, adaptive grids \cite{journals/cgf/Szirmay-KalosTM11}, and manually-graded tetrahedral mesh \cite{Fang:10} have been used to facilitate ray marching and/or sampling of scattering events through the media.

In volume rendering for scientific visualization, polyhedral meshes are commonly used with either ray casting \cite{Marmitt2006,wald:07:Tetty,muigg-2011-gpg,journals/vc/MirandaF12} or particle tracing \cite{conf/egpgv/BusslerRKHK11}. Polyhedral meshes provide smooth interpolation of the underlying volumetric field \cite{Wald05fasterisosurface} with its continuous structure, in contrast to structures like octrees that can have neighboring cells with different resolutions. Unstructured polyhedral mesh also provides the flexibility of adaptive cell sizes, which can either be constructed using a global scheme \cite{AtomicMeshing} that varies cell sizes in the entire mesh, or can be built dynamically using a top-down or bottom-up approach, resulting in a multi-resolution representation \cite{646238,Cignoni:1994:MMV:197938.197952}. Our algorithm uses a global approach similar to \cite{AtomicMeshing} to construct the tetrahedral mesh as a pre-process before ray traversal, while the latter methods can be useful for modeling dynamic media. A key difference between our approach and the methods proposed in the context of visualization is that, although the underlying volumetric function often represents density or other physical properties similar to our media profiles, volumetric ray casting generally does not account for the non-linear refractive paths that the light follows.

In the separate context of meteorology and Earth circulation modeling, unstructured mesh is advantageous due to its adaptive nature and its flexibility in terms of handling irregular domains. Consequently unstructured meshes have been increasingly adopted to replace regular grids in more recent operational models \cite{citeulike:10745077,dowling2013earth}. Models like \cite{citeulike:10745077,dowling2013earth} compute atmospheric flow fields at high resolution, which can provide detailed media profiles to serve as initial conditions for propagation. Therefore, adopting the unstructured mesh in propagation algorithms opens the possibility of seamless coupling between the atmospheric flow model and subsequent propagation within the resulting flow field.
\subsection{Analytic trajectories}
\label{review:analytic}
Analytic light paths have been derived in the context of geometric optics for simple profiles of refractive index \cite{ob:bornwolf,ob:ghatak,Kravtsov}. Cao et al.~\shortcite{Cao2010} is perhaps the first work in visual rendering to use the analytic ray formulation for constant gradient of the refractive index, based on the derivation in Qiao \shortcite{Qiao1984}. Cao et al. \cite{Cao2010} demonstrated the performance advantage over piecewise linear ray tracing, and used octrees for further acceleration. However, their ray formulation does not have a closed-form solution for intersections with planar surfaces; instead they used bisection methods.


Analytic rays with a polynomial formulation is proposed in \shortcite{Kerr2010} for artist-controlled lighting with curved rays. The light paths are not physically-based and cannot be easily extended to more realistic kinds of light bending from refraction. Grave et al.~\cite{grave-eurovis09} visualize the effects of general relativity using an analytic solution derived for the G\"{o}del universe.

In computational acoustics, closed-form ray trajectories have been derived for constant gradient condition in the propagation speed, and for constant gradient condition in the squared refractive index. The term {\it cell method} refers to acoustic ray tracing that subdivides media into cells and assumes closed-form ray paths in each cell, but it has only been used for 2D varying media modeled by regular triangular grid with no obstacles \cite{trimain,raywave}. Our algorithm can be seen as an extension of cell methods to a more general propagation algorithm that can handle 3D varying media and complex scene objects. Furthermore, we improve the efficiency based on closed-form ray intersections and use of an adaptive unstructured mesh.

\nocite{Qiao1984}
\nocite{Kerr2010}
\nocite{GSMA06}

\section{Background}
\label{section3}
\setlength{\belowdisplayskip}{-0.8pt}
\setlength{\abovedisplayskip}{-0.8pt}
In this section, we present background material on non-linear media and how it affects light and sound propagation.
\subsection{Non-linear media properties}
\label{raycurves-media}
The two most prominent non-linear media in outdoor scenes: the atmosphere and the ocean, are often studied separately. They are in fact tightly connected by heat flow and general circulation of the water component \cite{dowling2013earth}. We hereby focus our discussion on atmospheric properties, but we would like to point out that media properties and propagation in the ocean are analogous.

A standard profile of atmospheric temperature and pressure is available with the 1976 USA Standard Atmosphere \cite{usgpc1976standard}, which is a simple layered model based on averaged empirical measurements. On the other hand, atmospheric properties at any particular time and location deviate from the above Standard under different conditions. For example, atmospheric temperature has diurnal and seasonal variations, is affected by short-term weather patterns, and can fluctuate from heat sources nearby including human constructions and activities in an urban setting. The destandardized media profiles can be obtained from measurements, empirical models, or detailed simulations of atmospheric flow.

\subsubsection{Properties affecting light propagation}
\label{raycurves-light-properties}
Light propagation paths are governed by the spatial profile of refractive index, which can in turn be computed from atmospheric density and wavelength of the light. 

Starting from an atmospheric profile for a spatial location $\mathbf{x}$, density is computed from temperature and pressure using the Perfect Gas Law:
\begin{equation}
\rho(\mathbf{x})=\frac{P(\mathbf{x})M}{RT(\mathbf{x})},
\end{equation}
where $T$ is temperature, $P$ is pressure, $M$ and $R$ are constants with typical values of $28.96 \times 10^{-3} kg/mol$ and $8.3145 J/mol\cdot K$ respectively. The Cauchy's formula \cite{ob:bornwolf} relates index of refraction with wavelength as: $n(\lambda)=a\cdot{(1+\frac{b}{\lambda^2})}+1$, where $a$ and $b$ are constants with typical values of $a=2879 \times 10^{-5}$ and $b=567 \times 10^{-5}$ for air. The Gladstone-Dale Law \cite{dale1858influence} then represents $n(\lambda,\mathbf{x})$ as a function of both density $\rho(\mathbf{x})$ and $n(\lambda)$: $n(\mathbf{x},\lambda)=\rho(\mathbf{x})(n(\lambda)-1)+1$.


\subsubsection{Properties affecting sound propagation}
\label{raycurves-sound-properties}
The atmospheric speed of sound is governed by the temperature as 
\begin{equation}
\label{eq:sound-speed-temp}
c=\sqrt{\gamma R_d T_v},
\end{equation}
where $\gamma={c_p}/{c_v}$ is the ratio of the specific heats, $R_d$ is the gas constant of dry air, $T_v$ is the virtual temperature considering humidity, and can typically be approximated by the absolute temperature $T$ when the humidity effects are ignored. 

For sound propagation, the wind profile plays a role that is as important as the temperature. Within the surface layer close to the ground, a common wind profile based on the Monin-Obukhov similarity theory \cite{monin1957basic} computes the mean wind velocity as following a logarithmic law depending on the height. The same theory prescribes wind profiles for altitude beyond the surface layer with parameters representing stable and unstable atmospheric conditions \cite{panofsky1984atmospheric}. The wind profile can be incorporated into the {\it effective sound speed profile} by combining the temperature-based sound speed and the wind velocity \cite{l1993sound,lamancusa1993ray}.

The above profile requires detailed measured data for a particular location, time, and atmospheric condition. Alternatively, we can generate a representative media profile from physically-based empirical models of the atmosphere \cite{salomons2001}. The acoustic index of refraction in the atmosphere $n=c_0/c$, where $c_0$ is the reference sound speed, is modeled with a stratified component $n_{str}$ and a fluctuation component $n_{flu}$, so that $n=n_{str} + n_{flu}$.
The stratified component follows a logarithmic profile of the altitude $z$:
\begin{equation}
\label{eq:sound-profile-stratified}
	n_{str}(z) = c_0/(c_0 + b \ln\left(\frac{z}{z_g}+1\right)),
\end{equation}
with parameters $n_0$, $b$, and $z_g$. $c_0$ is the sound speed at the ground surface, taken as the reference sound speed, and $z_g$ is the roughness length of the ground surface. Typical values for parameter $b$ are $1$ m/s for a downward-refracting atmosphere and $-1$ m/s for an upward-refracting atmosphere. 

The fluctuation component models the random temperature and wind speed turbulence in the atmosphere. The value at position $\mathbf{x}$ can be computed as
\begin{equation}
\label{eq:sound-profile-plus-flux}
	n_{flu}(\mathbf{x})= \sum_{i} G(\mathbf{k}_i) 
		\cos( \mathbf{k}_i \cdot \mathbf{x} + \varphi_i),
\end{equation}
where $\mathbf{k}_i$ is the wave vector describing the \emph{spatial} frequency of the fluctuation, $\varphi_i$ is a random angle between $[0, 2\pi]$, and $G(\mathbf{k_i})$ is a normalization factor. The stratified-plus-fluctuation model is widely used in atmospheric acoustics~\cite{salomons2001}, and we adopted this model to generate realistic atmospheric benchmarks for our acoustic propagation in Section \ref{section6}.
\begin{figure*}
\centering
\setlength\fboxsep{0pt}
\setlength\fboxrule{1pt}
\fbox{\includegraphics[width=0.3\linewidth]{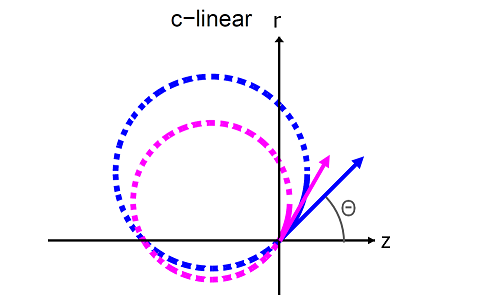}}
\fbox{\includegraphics[width=0.3\linewidth]{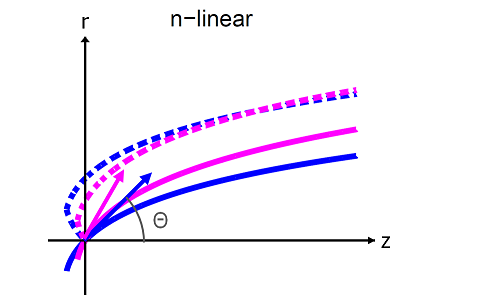}}
\fbox{\includegraphics[width=0.3\linewidth]{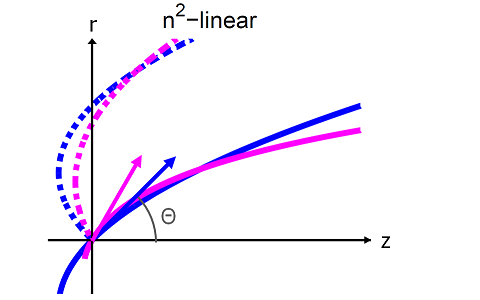}}
\caption{\textbf{Analytic ray curves} in $c$-linear (circular curves), $n$-linear, and $n^2$-linear (parabolic curves) media profiles. Our algorithm uses the circular and the parabolic curves as ray tracing primitives. Red and blue curves represent different launch angles (60\textdegree and 45\textdegree,  respectively). The dashed curves trace out the ray paths if the launch angles are flipped around the $r$-axis. The $z$-axis represents the direction of the media gradient. The $r$-$z$ plane is the \textit{ray plane} defined in Section \ref{section:ray-trajectories}.}
\label{ray_curves}
\end{figure*}

\subsection{Propagation trajectory}
\label{section:ray-trajectories}
In ray tracing for wave propagation, rays are defined as normal to the wavefront. The Eikonal equation for ray trajectories is derived from the wave equation as follows:
\begin{subequations}
\begin{equation}
\label{eq:ray-eq-c}
\frac{d}{ds}\left(\frac{1}{c(\mathbf{x})}\frac{d\mathbf{x}}{ds}\right)=-\frac{1}{c(\mathbf{x})^2}\nabla c(\mathbf{x}),
\end{equation}
\begin{equation}
\label{eq:ray-eq-n2}
\frac{d}{ds}\left(n(\mathbf{x})\frac{d\mathbf{x}}{ds}\right)=\nabla n(\mathbf{x}),
\end{equation}
\end{subequations}

\noindent where $\mathbf{x}=\{x,y,z\}$ represents the Cartesian coordinates, $s$ is the arc-length along the ray, $c(\mathbf{x})$ is the propagation speed that is a function of the spatial location, $n(\mathbf{x})=c_0/c(\mathbf{x})$ is the index of refraction, and $c_0$ is the reference propagation speed. 

Real-world media such as the atmosphere tend to vary smoothly and therefore can be modeled with continuous functions with locally varying gradients. Given a local media gradient at location $\mathbf{x}$, we hereby provide the analytic ray trajectory in a local coordinate system aligned with the gradient direction. 

For a particular ray origin $\mathbf{x}$ and direction $\mathbf{d}$, we place the origin of the local coordinate system at $\mathbf{x}$, and denote the media gradient direction as the $z$-axis. It can be shown that the ray trajectory is a plane curve that lies in the plane formed by the gradient direction and the ray direction $\mathbf{d}$, i.e. the {\it ray plane}. We then select the direction within the ray plane that's perpendicular to the $z$-axis as the $r$-axis. Figure \ref{ray_curves} plots the analytic ray curves in the \textit{ray plane}. 

For the local gradient in propagation speed $c$, $\alpha=\|\nabla c\|$, the local profile can be written as: $c(z)=c_0+\alpha z$, where $c_0$ is $c$ at the ray origin. Let $\xi_0'=\frac{cos\theta_0}{c_0}$, where $\theta_0$ is the angle between initial ray direction and the $r$ axis, the ray trajectory in $r$-$z$ coordinates can be derived from Equation (\ref{eq:ray-eq-c}) to be:
\begin{equation}
\label{eq:c-linear-r-front}
r(z)=\frac{\sqrt{1-\xi_{0}'^{2}c_0^2}-\sqrt{1-\xi_{0}'^{2}\left(c_0+\alpha z\right)^2}}{\xi_0'\alpha},
\end{equation}
which is a circular curve in the ray plane. (See Appendix \ref{appendix-derivation} for detailed derivations.)

For the local gradient in refractive index $n$, the analytic ray curve for $\nabla n$ was derived in \cite{Cao2010}, which does not have an analytic solution for intersection tests. For the local gradient $\nabla n^2$, however, an analytic ray trajectory with analytic intersection solution exists. We establish a similar coordinate system with origin at a spatial location $\mathbf{x}$, and $z$-axis parallel with the local media gradient $\nabla n^2$. 

For $\alpha=\|\nabla n^2\|$, the local profile can be written as: $n^2(z)=n_0^2+\alpha z$, where $n_0$ is the value of $n$ at the ray origin. Let $\xi_0'=n_0 cos\theta_0$, where $\theta_0$ is the angle between initial ray direction and the $r$ axis, the ray trajectory is:
\begin{equation}
\label{eq:n2-linear-r-front}
r(z)=\frac{2\xi_0'}{\alpha}\left(\sqrt{-\xi_{0}'^{2}+n_0^2+\alpha z}-\sqrt{-\xi_{0}'^{2}+n_0^2}\right),
\end{equation}
\noindent which is a parabolic curve. (See Appendix \ref{appendix-derivation} for detailed derivations.)

Both the circular and the parabolic ray curves have closed-form solutions in terms of intersections with planar surfaces, and for travel distance along the ray. We have plotted the $n$-linear ray curve used by Cao et al. \cite{Cao2010}, and the $c$-linear and $n^2$-linear ray curves in Figure \ref{ray_curves}.


\section{Adaptive Media Mesh}
\label{section5}

\begin{figure*}
\centering
\subfloat{\includegraphics[width=2.0\columnwidth]{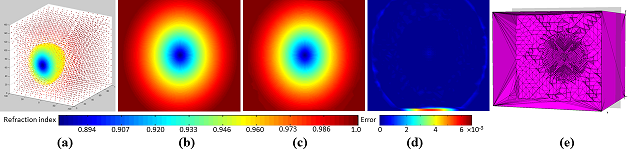}}
\\
\subfloat{\includegraphics[width=2.0\columnwidth]{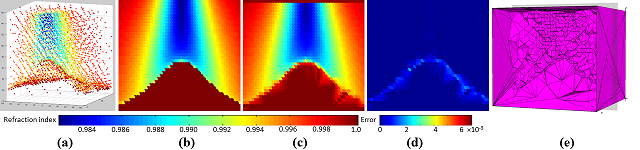}}
\caption{\textbf{Adaptive meshes.} The unstructured meshes we construct (Section \ref{section5}) have the capability to adapt to complex spatial media profiles. Here we show two meshes generated for the hot spot (\textbf{A-HS}) and upwind-over-hill (\textbf{A-UW}) profiles, respectively. \textbf{(a)} Resampled media points (showing half of the points to expose the sectional view), \textbf{(b)} Input media profiles, \textbf{(c)} Interpolated media profiles from the meshes, \textbf{(d)} Absolute approximation errors, \textbf{(e)} Adaptive meshes. The input media grid has $6.4\times10^6$ ($200\times200\times160$) points for A-HS, and $8\times10^5$ ($100\times100\times80$) points for A-UW. The meshes are constructed from a resampled $4.3\times10^4$ points for A-HS, and $9.8\times10^3$ points for A-UW. With 100$\times$ fewer sample points than the input the adaptive meshes are able to achieve low approximation errors in the media profiles they represent.}
\label{result:adaptive_mesh}
\end{figure*}

Our goal is to construct a tetrahedral mesh with graded cell sizes that adapts to the spatial distribution of media properties, hereafter referred to as an \textit{adaptive mesh}. The cost of computing a ray curve and its intersection within each media cell is constant, therefore an adaptive mesh leads to faster ray traversal. We also take advantage of the capability of an unstructured tetrahedral mesh to conform to arbitrary surfaces, and we embed boundary surfaces that represent scene objects. There are multiple ways to construct an adaptive mesh and to incorporate boundary surfaces into it. In this section, we give details of the techniques used in our implementation and also discuss some alternatives. These discussions are substantiated with experimental results in Section~\ref{section6}.

\subsection{Resample media profiles}
\label{resample}
We assume that the input media profiles are available on a three-dimensional uniform grid. The data points on the grid are generated from real-world measurements or from sampling a characteristic profile. We will now describe our method of resampling an input profile to generate a point set distributed according to local magnitude of media variations; tetrahedralization on such a point set generates an efficient structure for both media representation and ray traversal.

We want to vary the cell sizes according to media variations. To achieve this goal, we want to vary the spacing between sample points when we resample the input media profile. For example, given an input profile of propagation speed $c(\mathbf{x})$ for each grid point at location $\mathbf{x}$, we can compute the \textit{slowness} $k(\mathbf{x})=\frac{1}{c(\mathbf{x})}$, and the gradient of the slowness $\nabla k(\mathbf{x})$ on the input grid by finite difference. We then compute a spacing $d(\mathbf{x})$ such that $\sigma=\frac{1}{4}\nabla k d^2(\mathbf{x})$, with a global $\sigma$ that controls the overall variation allowed in each cell.

After computing the desired spacing $d(\mathbf{x})$ for each grid point location $\mathbf{x}$ in the profile, we use Algorithm~\ref{construct} to obtain the set of resampled points $S$ from the profile $G$, in a manner similar to the \emph{Atomic Meshing} process~\cite{AtomicMeshing}. Basically, a \emph{face-centered-cubic} (FCC) lattice is grown from the center of the space outward, placing each new point away from existing samples by the spacing $d(\mathbf{x})$. The approximation errors that are introduced by the resampling process are quantified in Section \ref{section_error}.

\LinesNumbered
\begin{algorithm}[h]
\SetAlgoLined
Given the set of grid points of the media profile $G$, initialize a flag array that marks each point in $G$ as \emph{false}\;
Initialize an empty list of points for output $S$,
and a queue of points $T$ with just the center point in the grid $\mathbf{x}_i$ in it\;
\While{$T$ is not empty}{
dequeue $\mathbf{x}_i$\;
\If{$\mathbf{x}_i$ lies within the bounds of the profile}{
   compute a spherical region with center $\mathbf{x}_i$ and radius $d(\mathbf{x}_i)$\;
   \If{all samples in the spherical region are marked \emph{false}}{
   mark all such samples \emph{true}\;
   add $\mathbf{x}_i$ to $S$;
   compute ideal sites of $\mathbf{x}_i$ with spacing $d(\mathbf{x}_i)$ and enqueue all in $T$\;
   }
}
}
\vspace*{0.1in}
\caption{\textbf{Media sample redistribution for mesh construction.} We adjust spacing for a given input profile on a regular grid. The \emph{ideal sites} are the locations of neighbors in a FCC lattice \cite{AtomicMeshing}.}
\label{construct}
\vspace*{-0.07in}
\end{algorithm}

\subsection{Embed boundary surfaces}
\label{embed}
During propagation, objects in the scenes (such as terrains, mountains, man-made structures such as buildings and sound barriers) affect the propagation paths of the curved rays. Given the tetrahedral mesh used to represent the media, surfaces that represent the scene objects can be incorporated by either embedding them in the mesh or linking them to the mesh cells that they overlap. 

Unlike axis-aligned data structures, such as octrees, that are commonly used in ray tracing, tetrahedral mesh has flexible structures that can embed surfaces of arbitrary orientations. To embed surfaces, we insert them as boundary constraints and construct a constrained tetrahedral mesh. When the surfaces are embedded in the mesh, no separate intersections with surfaces are computed during the ray traversal, and Line~\ref{intersection} in Algorithm~\ref{traversalA} is merged with Line~\ref{intersect-face}. When a ray's exit face from a tetrahedral cell corresponds to a constrained face (an object's boundary surface), the current ray traversal terminates and a secondary ray is spawned reflecting off the constrained face. 

While embedding the boundary surfaces often brings speedup, because it unifies mesh traversal with surface intersections and eliminates extra computation during traversal, inserting constrained surfaces adds considerable computational overhead to the mesh construction. We evaluate this trade-off between construction efficiency and traversal efficiency individually for each input scene.

In our benchmarks, the distribution of surface primitives is always compatible with media variation, leading to the fast traversal of a constrained mesh. However, if there are over-tessellated objects or objects whose resolution doesn't match with the media variation, the constrained mesh generation algorithm chooses smaller cell sizes close to the objects' surfaces and this affects the traversal performance.

To compensate for scenarios like this, there are two options. We could treat the object boundaries as implicit surfaces instead of explicit triangles. In this case, we can construct a tetrahedral mesh that conforms to these implicit surfaces, effectively re-tessellating it, as in \cite{10.1109/TVCG.2013.115}. This method keeps the performance benefit of a unified traversal while maintaining proper cell size, i.e. the cell size based on media variations instead of object tessellations. Or we could link each tetrahedral cell to a list of the boundary faces that it overlaps with, similar to \cite{Cao2010}; in this case, the ray traversal of each cell needs to iterate through this list of boundary faces to compute the surface intersections. This approach has the benefit of simplicity, but might not provide optimal traversal performance. Furthermore, generating those links comes with its own computational overhead. We report the experimental results of the performance of both construction and traversal of the meshes when embedding or linking boundary surfaces (See Section \ref{others}).

\subsection{Tetrahedralization}
Given a point set $S$ on a lattice with proper spacing (extracted from the media profile as described in Section~\ref{resample}), and given the optional constrained surfaces of the objects ($P$) in the scene (discussed in Section~\ref{embed}), we use these two sets of constraints to compute a Constrained Delaunay Tetrahedralization (CDT).


We use the method proposed by Si and Gadrtner \shortcite{Si05} and implemented in the \emph{TetGen} software package, to build a CDT with $S$ and $P$. The resulting CDTs have adaptively graded cell sizes due to resampling of the input profile, and we observe well-shaped tetrahedral mesh with a maximum radius-edge ratio below 2.0 in our benchmarks. The CDT construction process can potentially insert additional points into the mesh; we obtain the inserted points' media properties by querying and interpolating the original input profile. The set of boundary surfaces ($P$) becomes constrained faces of the constructed CDT, which are stored as face markers with each tetrahedral cell in the mesh.

\nocite{Si05}

\section{Traversal of Ray Curves}
\label{section-mesh}

After we construct an adaptive tetrahedral mesh based on the input media profile, propagation through this media can be simulated by computing ray curves from media gradients estimated over the mesh and traversing those ray curves using the mesh connectivity.

\subsection{Gradient estimation}
\label{section:gradient}
Given the spatial decomposition of the media profile with our tetrahedral mesh, we need accurate estimation of the media gradient within each tetrahedral cell to compute the analytic ray trajectories entering that cell. Our method is based on the assumption that the per-cell local gradient captures the media variation within that cell, and we adopted a cell-centered linear regression-based gradient estimation method.

For media property $m$ (e.g. $c$ or $n^2$) defined over the domain, and a cell $C$ in the mesh with centroid $\mathbf{x}_0$, the cell gradient $\nabla m$ should satisfy the equation system:
\begin{equation}
\label{eq:regression}
\begin{bmatrix}
(\mathbf{x}_1-\mathbf{x}_0)^\intercal \\
(\mathbf{x}_2-\mathbf{x}_0)^\intercal \\
(\mathbf{x}_3-\mathbf{x}_0)^\intercal \\
(\mathbf{x}_4-\mathbf{x}_0)^\intercal
\end{bmatrix}\nabla m = 
\begin{bmatrix}
m(\mathbf{x}_1)-m(\mathbf{x}_0) \\
m(\mathbf{x}_2)-m(\mathbf{x}_0) \\
m(\mathbf{x}_3)-m(\mathbf{x}_0) \\
m(\mathbf{x}_4)-m(\mathbf{x}_0) \\
\end{bmatrix},
\end{equation} 
where $\mathbf{x}_k,k=1,...,4$ are the centroids of the 4 neighbors of $C$, $m(\mathbf{x}_k)$ is the media property values at those centroids. Written in matrix form:
\begin{equation}
\label{eq:regression_matrix}
\mathbf{X} \nabla m=\mathbf{b},
\end{equation}
Optionally, different weights can be assigned to each neighbor of the cell, to take into consideration of the irregular shapes of the mesh:
\begin{equation}
\label{eq:regression_matrix_weight}
\mathbf{W}\mathbf{X} \nabla m=\mathbf{W}\mathbf{b}
\end{equation}
where $\mathbf{W}=diag\{w_i\}$ is a $4\times4$ diagonal matrix containing the weights of neighbor $k$ of cell $C$. This can be solved with linear least square (See Appendix \ref{appendix-gradient} for the explicit solution of the estimated gradient).

Although average-based gradient estimation method is faster to compute, and has been used in prior work \cite{Cao2010}, the regression-based method, especially the weighted version with inverse centroid distance, has been shown to provide better accuracy for irregular shaped mesh elements, and adapts well to lower-quality meshes \cite{correa2011comparison,mavriplis2003revisiting}. Accuracy of the estimated gradient is particularly important for outdoor propagation, when artifacts such as false caustics have been shown to happen with discontinuous gradients \cite{jones1986harpa}. We estimate the gradient as a preprocess right after adaptive mesh construction, and we compare the results with the Green-Gauss method used by \cite{Cao2010}, which is essentially volume-weighted averaging-based method (Section \ref{results:comparison_cao}). 
\subsection{Curved ray traversal}
\label{section5-1} 
The pseudo-code for the traversal of curved rays through a tetrahedral mesh is given in Algorithm~\ref{traversalA}. 

Given a ray origin, we first locate the tetrahedral cell that contains the origin. This step is commonly referred to as \emph{point location}, and it can be relatively expensive for complex models when there are a large number of tetrahedral cells. However, in most scenarios, each primary ray originates from the same point (light or sound source), and each secondary ray (after interacting with boundary surfaces) originates from the same cell where its predecessor (the primary ray that spawned it) ends. The point-location query is performed once per frame, and the cost is amortized over all the rays.
 
\LinesNumbered
\begin{algorithm}[h]
\SetAlgoLined
Point Location for ray origin $P$, yields tetrahedron $T$\;
Compute analytic ray trajectory from media property of $T$\;\label{next}
Intersect ray curve with $T$ to find exit face $F$\;\label{intersect-face}
\uIf{$T$ contains boundary surfaces}{
   surface interaction\;
   go to \ref{next} with $T$ unchanged\;
   }\label{intersection}
\uElseIf{there is a tetrahedron $T'$ incident to $F$}
   {$T=T'$; Go to Step \ref{next}\;}\label{neighbor-move}
\Else{ray exits the scene\;}
\vspace*{0.1in}
\caption{\textbf{Curved Ray Traversal} of tetrahedral meshes.}
\label{traversalA}
\vspace*{-0.10in}
\end{algorithm}

Once the initial tetrahedron is located, we retrieve the interpolated media properties $\nabla m$, which have been precomputed and stored in the tetrahedral cell.  The direction of $\nabla m$ and the initial location and direction of a ray are used to define the {\it ray plane}, and we can compute the curved trajectory within the cell for any entering ray, as described in Equation~\ref{eq:c-linear-r} or \ref{eq:n2-linear-r} (Line~\ref{next} in Algorithm \ref{traversalA}).

The ray curves we used have closed-form intersection solutions with planar surfaces, e.g. the four faces of the tetrahedral cell. The intersection point closest to the ray origin is chosen as the exit point from the cell, and the neighboring cell incident to the exit face is taken as the next cell in the traversal. We use its media properties to compute the next segment of the curved ray path (Line~\ref{neighbor-move} in Algorithm \ref{traversalA}).
\subsection{Boundary and media interactions}
\label{section-boundary}
Unstructured tetrahedral mesh has the capability to conform to irregular boundary surfaces, and we choose to leverage this capability to embed surfaces in the mesh, as mentioned in Section \ref{embed}. In this case, the ray encounters boundary faces when it traverses those tetrahedra in which the boundary surfaces are embedded. 

Given the ray trajectories in Equation~\ref{eq:c-linear-r-front} and \ref{eq:n2-linear-r-front}, the tangent direction of the ray at arbitrary point (e.g. an intersection point) along the ray curve can be evaluated analytically for the circular curve:
\begin{equation}
\frac{dr}{dz}=\frac{(\xi_0'(\alpha z+c_0))}{\sqrt{1-\xi_0'^2 (\alpha x+c_0)^2}},
\end{equation}
and for the parabolic curve:
\begin{equation}
\frac{dr}{dz}=\frac{\xi_0'}{\sqrt{-\xi_0'^2+\alpha x+n_0^2}}.
\end{equation}

With the incident location and direction of the ray at a boundary surface, perfect reflection, Snell's law refraction, or BRDF-based sampling can be employed to generate the direction of the next ray along the propagation path. The new ray will be computed based on the media property of either the current cell (for reflecting surfaces) or the neighboring cell (for the refracting surfaces) and the new direction (Line~\ref{intersection} in Algorithm \ref{traversalA}). 

Furthermore, the circular and parabolic ray curves both have closed-form arc length, which can be used to compute attenuation of propagated energy due to absorption for light (e.g. \cite{Ament:2014:RRT:2603314.2557605}) and sound (\cite{jensen2011}). A closed-form arc length is also convenient for free path sampling to simulate media scattering \cite{Yue:2010,journals/cgf/Szirmay-KalosTM11}. While our ray formulation is compatible with more complex surface interactions and media participation, we do not perform BRDF sampling or media scattering in our benchmark results but focus on the refractive characteristics of non-linear media and specular boundary reflections.


\section{Results and Analysis}
\label{section6}

\begin{figure}
\begin{center}
\subfloat[]{\includegraphics[width=0.45\columnwidth]{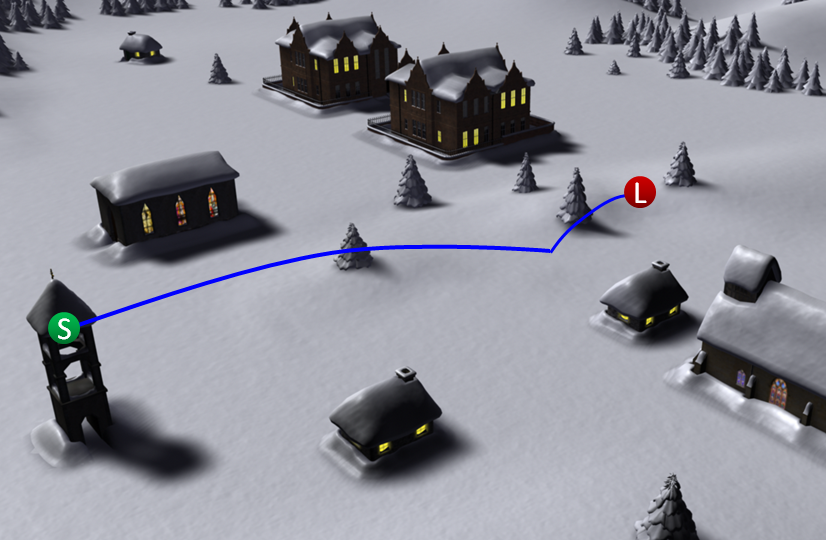}}
\quad
\subfloat[]{\includegraphics[width=0.45\columnwidth]{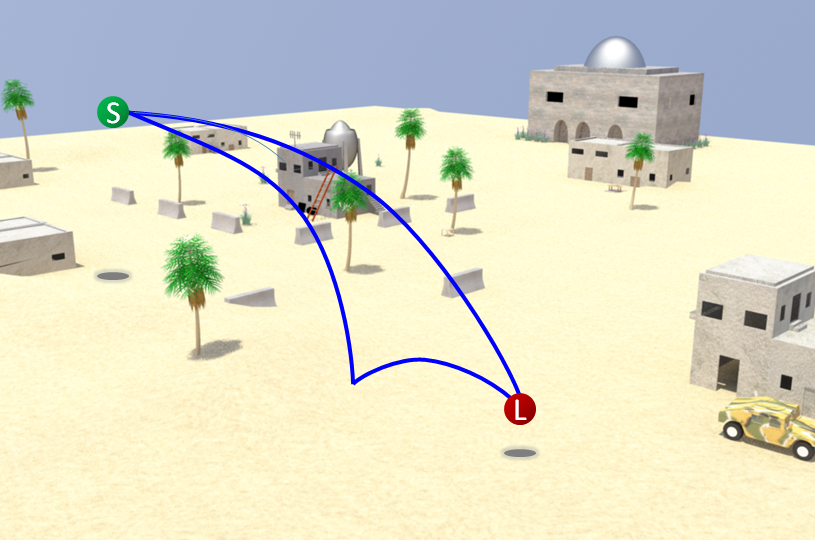}}
\caption{\textbf{Acoustic benchmarks} The Desert and Christmas shown here represent complex outdoor acoustic scenes (details in Table \ref{table:scenes}). Propagation results for these benchmarks can be seen in Figure \ref{result:3dray} (in the Appendix), Figure \ref{media_scale}, and Table \ref{table:breakdown}.}
\label{test-scenes}
\end{center}
\end{figure}

In this section, we highlight the applications of our algorithm on light and sound propagation in outdoor benchmarks with different atmospheric profiles and geometric primitives in the scene. The efficiency of the curved ray tracer enables simulation of general media with 3D variations interacting with complex boundaries, which have not been achieved before. We compare the performance with piece-wise linear ray stepping, widely used for non-linear media, as well as \cite{Cao2010}, the previous work in computer graphics that traces analytic ray curves. We also analyze the approximation errors incurred by the adaptive mesh we construct to represent the underlying media. The other implementation choices  including boundary surface embedding and regression-based gradient estimation are also analyzed with experimental results. 
\subsection{Benchmarks}
\label{section6-1}
Propagation in non-linear media is important for both visual and acoustic applications, therefore we tested our algorithm on visual benchmarks (Figure \ref{render_cao} and \ref{result:mirage} in the Appendix) and acoustic benchmarks (Figure \ref{result:2dray}, \ref{test-scenes}, and Figure \ref{result:3dray} in the Appendix).
 Each of the benchmarks consists of two components: a media profile, and triangulated geometric representation of the boundary surfaces.

\subsubsection{Media profiles} We generate media profiles that resemble realistic atmosphere under a set of different conditions, and we use different techniques for light and sound propagation. Some of these profiles have been used by previous work to model the atmosphere.

The profiles used for visual benchmarks include:
\begin{itemize}
\item \textbf{Inferior mirage (V-IM)}(modeled in \cite{khular1977note}), with the squared refractive index:
\begin{equation}
n^2(z)=\mu_0^2+\mu_1^2(1-exp(-\beta z)),
\end{equation}
where $z$ is the height, $\mu_0=1.000233,\mu_1=0.4584,\beta=2.303$.
\item \textbf{Superior mirage (V-SM)} represents atmospheric condition that is amenable to superior mirages (modeled in \cite{khular1977note}):
\begin{equation}
n^2(z)=\mu_0^2+\mu_1^2exp(-\beta z),
\end{equation}
These two profiles are also used in \cite{Cao2010}.
\end{itemize}
The refractive index profile for light waves is generated using the method described in Section \ref{raycurves-light-properties}.

The profiles used for acoustic benchmarks include:
\begin{itemize}
\item \textbf{Upward or downward refractive stratified profile (A-LU,A-LD)} represents the profile defined by Equation \ref{eq:sound-profile-stratified} as described in Section \ref{raycurves-sound-properties}. We use $n_0$ = 1, $c_0$ = 340 m/s, and $z_g$ = 1 m. We take $b=1 m/s$ for A-LD and $b=-1 m/s$ for A-LU.
\item \textbf{Hot spot (A-HS)} represents the localized heat source induced sound speed fluctuations, computed according to Equation \ref{eq:sound-speed-temp} and Equation \ref{eq:hotspot}, superimposed on an upward refractive stratified atmosphere profile (A-LU).
\begin{equation}
\label{eq:hotspot}
T=T_0+(T_s-T_0)exp(-d/d_0),
\end{equation}
where $T_s$ is the temperature at the hot spot, $d$ is the distance to the hot spot, $T_0=273K$ and $d_0$ is the dropoff length, which is a variable property of the hot spot. 
\item \textbf{Stratified-plus-fluctuation (A-LU+F, A-LD+F)} represents an upward or downward refractive atmosphere (A-LU or A-LD) superimposed with fluctuations generated according to Equation \ref{eq:sound-profile-plus-flux} and Section \ref{raycurves-sound-properties}.
\item \textbf{Wind over hill (A-UW for upwind, A-DW for downwind)} represents a known wind profile over an analytic hill shape \cite{jackson1975turbulent}. According to the Monin-Obukhov similarity theory \cite{monin1957basic}, the mean wind velocity follows the logarithmic law with height $z$: 
\begin{equation}
u(z)=\frac{u_\ast}{K}\ln{\frac{z}{z_g}},
\end{equation}
where $K$ is the von-Karmann constant, $z_g$ is the aerodynamic roughness length, and $u_\ast$ is the friction velocity \cite{businger1971flux,oke1987boundary}. 

Above undulating terrains, the wind profile will be significantly modified based on the shape and properties of the ground. In particular, Jackson and Hunt \cite{jackson1975turbulent} derived closed form solution for a hill of the shape:
\begin{equation}
f(\frac{x}{L})=\frac{1}{1+(\frac{x}{L}^2)},
\end{equation}
where $x$ is the horizontal distance of a location from the apex of the hill, $L$ is the radius of the base of the hill. The analytic solution for the horizontal component of the wind velocity over this particular hill shape, in addition to the mean velocity $u(z)$, is given as:
\begin{multline}
\Delta u=u_0(z=L)\frac{h}{L}\frac{\ln(\frac{L}{z_0})}{\ln^2(\frac{l}{z_0})}(\frac{1-(\frac{x}{L})^2}{1+(\frac{x}{L})^2}\ln(\frac{\Delta z}{z_0})\\
-(\frac{2(x/L)}{(1+(x/L)^2)^2}(\frac{\Delta z-z_0}{l})\ln(\frac{\Delta z}{z_0})),
\end{multline}
where $\delta z$ is the distance above the hill, and $l$ is the thickness of the hill's influence region, where the flow above the ground is perturbed by the presence of the hill. $u(z)+\Delta u$ is then added to or subtracted from the underlying sound speed, for upwind or downwind propagation respectively, to form an effective sound speed profile \cite{salomons2001}.
\end{itemize}

\subsubsection{Geometric models}
The Desert and Christmas models (See Figure \ref{test-scenes}) represent large-volume outdoor acoustic scenes that have complex surface geometry (e.g. varying terrains and buildings). The details for these models are given in Table \ref{table:scenes}, including the surface primitive count, the number of re-sampled media points, and the size of the adaptive mesh constructed using Algorithm \ref{construct}.
\begin{table}[ht]
\centering
\begin{tabular}{c c c c c}
\hline\hline
Scene & \# surfaces & \# medium points. & \# tetrahedra \\[0.5ex]
\hline
Elephant & 1,500 & 1,532 & 5,538 \\
Desert (m) & 8,000 & 23,632 & 144,976 \\
Desert (h) & 16,000 & 132,742 & 674,434 \\
Christmas (m) & 8,000 & 44,862 & 227,851 \\
Christmas (h) & 16,000 & 179,382 & 1,169,353 \\
\hline
\end{tabular}
\caption{Benchmarks details. Acoustic benchmarks are tested at a range of different resolutions, and we show the stats at the median(m) and high(h) ends of the range.}
\label{table:scenes}
\end{table}
\subsection{Performance of curved ray traversal}
\label{section6-3}
\begin{figure*}
\centering
\subfloat[]{\includegraphics[height=1.8in]{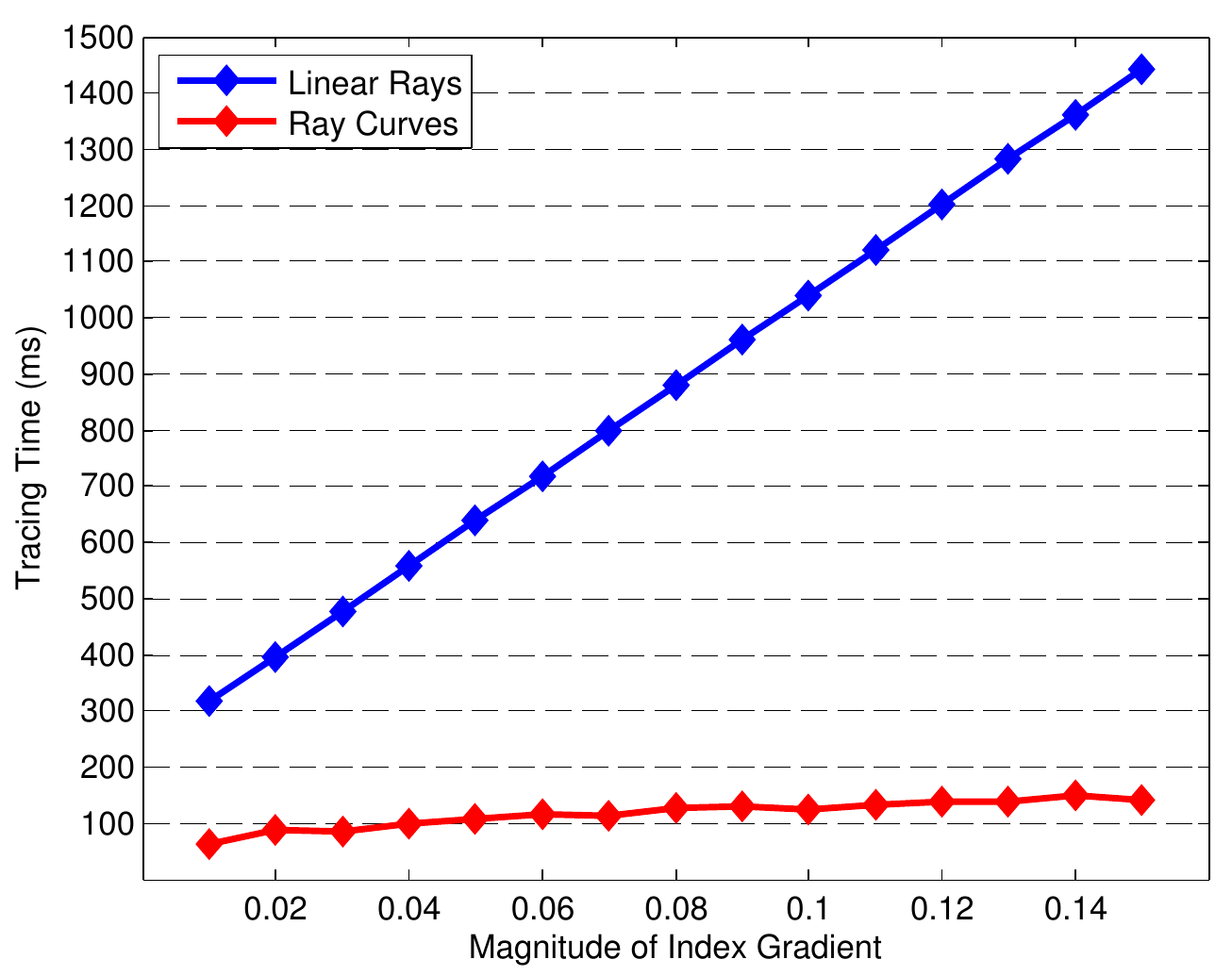}
\label{media_scale:a}}
\subfloat[]{\includegraphics[height=1.8in]{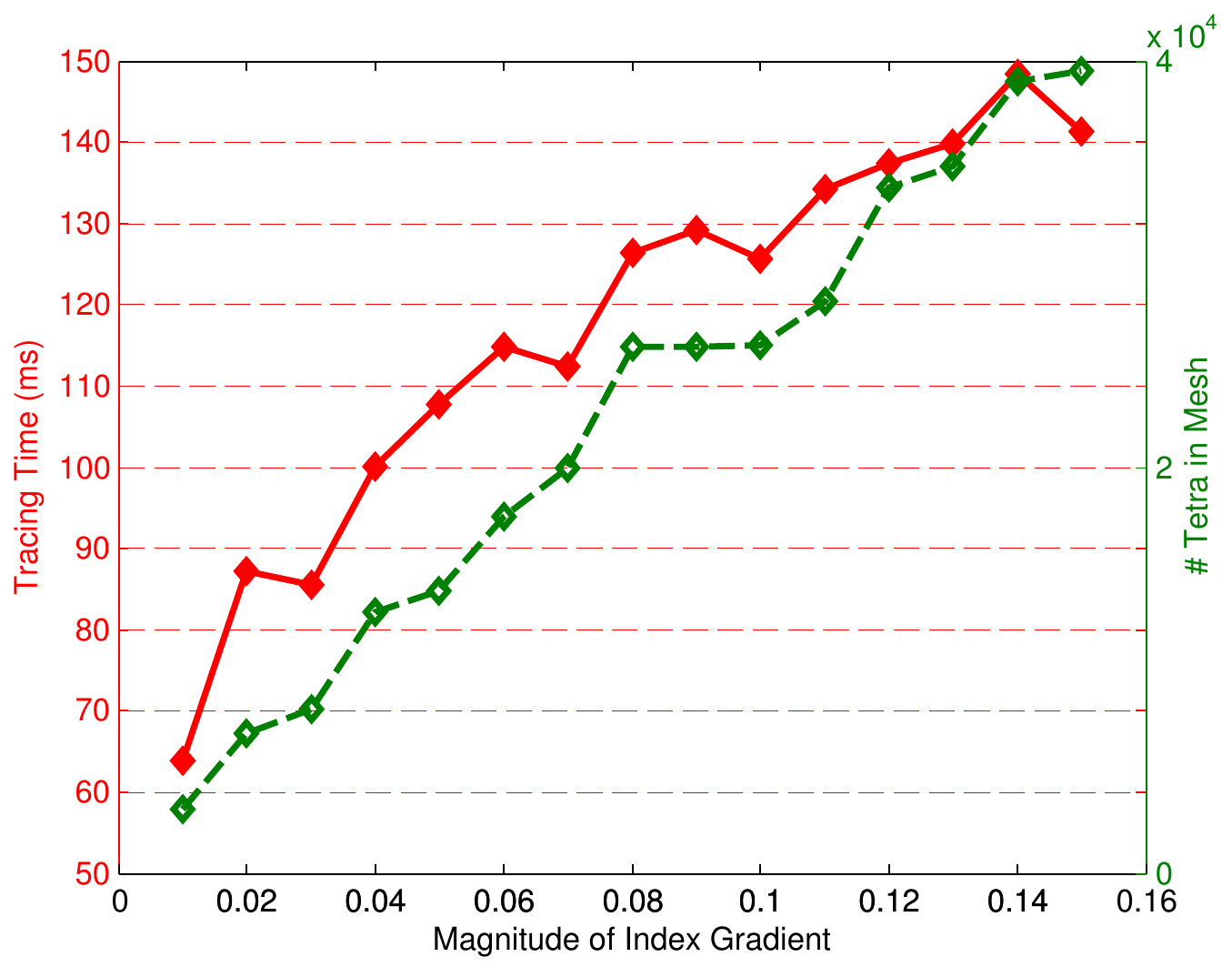}
\label{media_scale:b}}
\\
\subfloat[]{\includegraphics[height=1.8in]{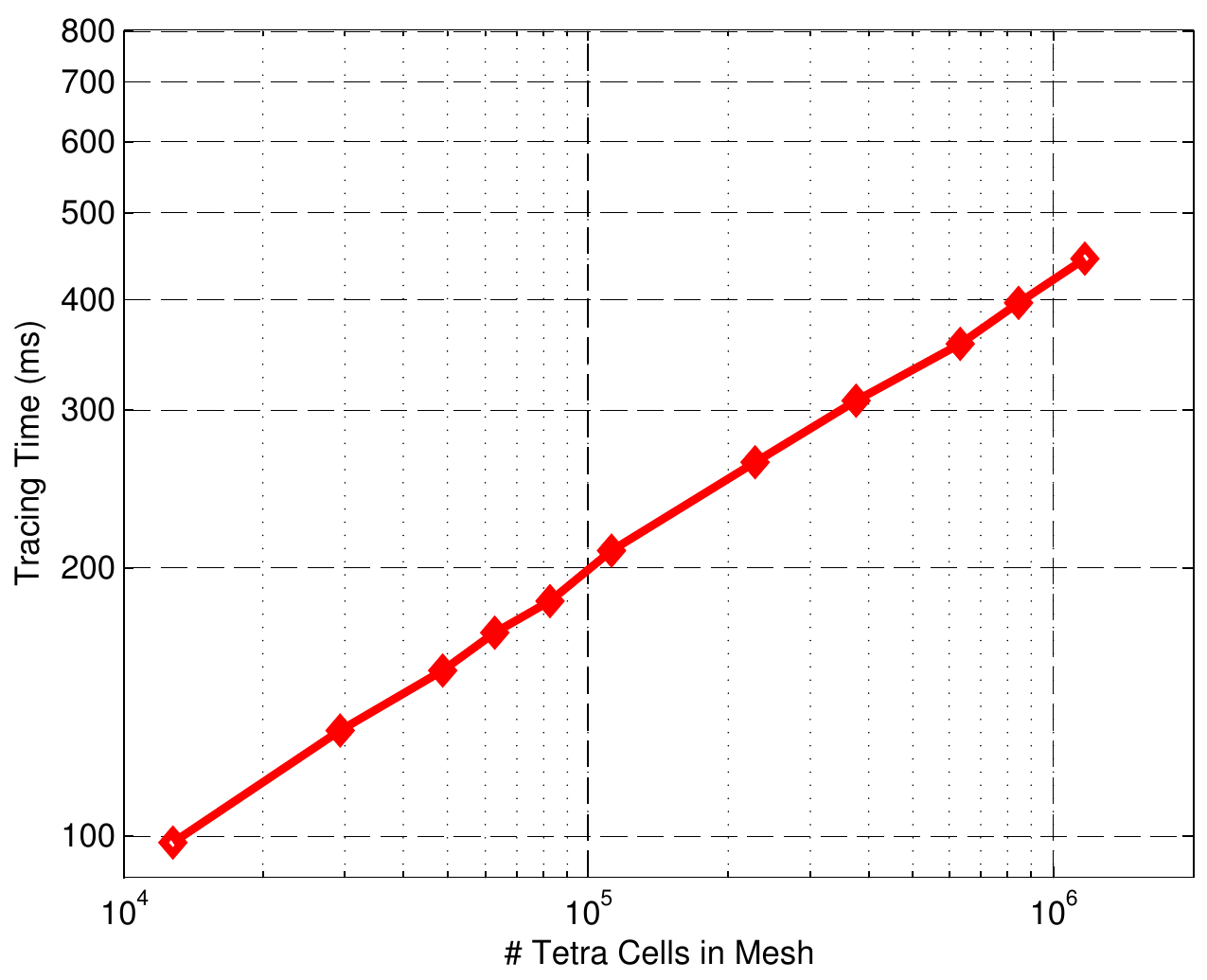}
\label{media_scale:c}}
\subfloat[]{\includegraphics[height=1.8in]{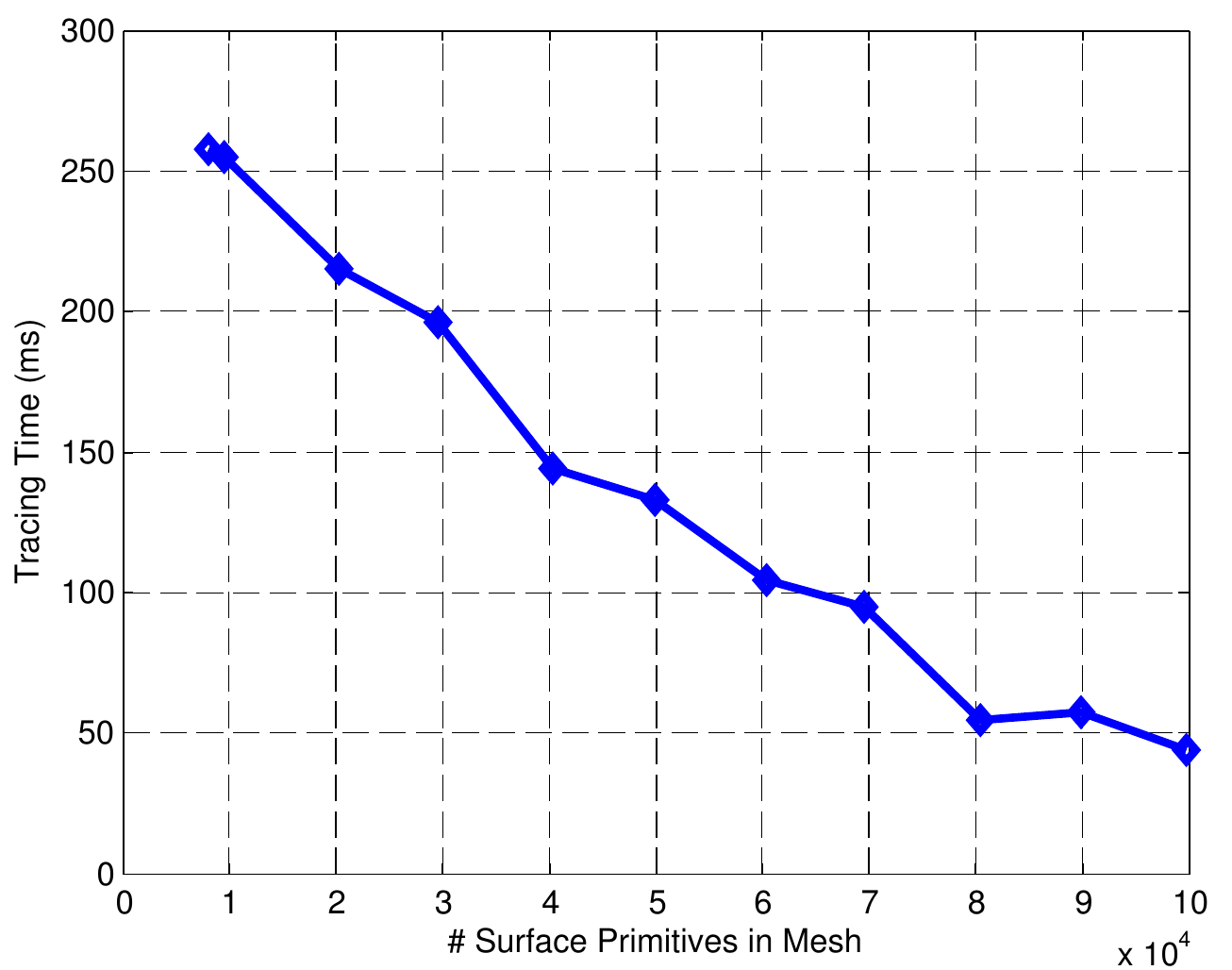}
\label{media_scale:d}}
\caption{\textbf{Performance and scalability of curved ray tracing}: \textbf{(a)} Tracing analytic ray curves vs. tracing linear ray segments, when simulating a sound propagation path to the same accuracy. Curved ray tracing scaled much better with increasing magnitude of media variations. \textbf{(b)} A close-up view of the "Ray Curves" line plot in (a) shows the tracing time (red line) scaling with increasing mesh sizes (green line). The increasing mesh size is a result of the adaptive mesh construction that keeps the approximation error at the same level, using a  lager number of smaller cells for increased media gradient. \textbf{(c,d)} Curved ray tracing scales sub-linearly with tetrahedral cell counts and number of boundary surfaces in the mesh. Note that tracing time decreases with increased number of surface primitives because the average propagation distance before a ray bounces off a boundary surface is shortened.}
\label{media_scale}
\vspace*{-0.15in}
\end{figure*}

\begin{table*}[ht]
\centering
\begin{tabular}{c|c|c c c|c}
\hline\hline
Benchmark & Frame time & Compute Curves & Tetra Intersect (time) & Tetra Intersect (count) & Bisection\\[0.5ex]
\hline
Elephant & 123 & 0.0175 (0.01\%) & 110.08 (88.06\%) & 51 & +108.49\\
Desert (m) & 219 & 0.0658 (0.03\%) & 211.39 (96.24\%) & 179 & +247.70\\
Desert (h) & 369 & 0.1033 (0.03\%) & 361.59 (97.92\%) & 254 & +443.12\\
Christmas (m) & 259 & 0.1037 (0.04\%) & 240.96 (92.89\%) & 188 & +220.98\\
Christmas (h) & 443 & 0.1948 (0.04\%) & 427.99 (96.64\%) & 296 & +392.73\\[1ex]
\hline
\end{tabular}
\caption{\textbf{Breakdown of curved-ray traversal time}, tracing $10$K rays to a depth of 3. Tetrahedral cell intersection dominates the frame time, while ray curve formulation and computation cost is negligible. We also report the average number of tetrahedra that each ray curve traverses. For comparison with \cite{Cao2010}, we trace $n$-linear rays (shown in Figure \ref{ray_curves} (middle)) for the same scene configurations, and report the additional time that bisection takes in the rightmost column. Our ray formulations avoids the bisection computation due to their analytic surface intersections. All timings are in milliseconds.}
\label{table:breakdown}
\end{table*}

We show the performance advantage of tracing analytic curved rays over tracing piece-wise linear ray steps for visual rendering in an outdoor atmospheric environment (Figure \ref{result:mirage} in the Appendix). All the timings are collected on a single 3.2GHz CPU core. Under the V-IM and V-SM profiles, we performed same-quality comparison by carefully adjusting the step size of piece-wise linear ray tracing to match the trajectory produced by curved ray tracing. The performance of curved ray tracer is an order of magnitude faster under the same-quality comparison, while the piece-wise linear rays lead to noticeable artifacts when running at competitive speed (same-speed comparison).

We analyze the performance for sound propagation in greater detail (Figure~\ref{media_scale}), as the advantage of curved ray tracing in that context, unlike visual rendering \cite{Cao2010}, has not been sufficiently explored. We observe significant performance improvement and better scalability with curved ray tracing . In contrast, piece-wise linear ray stepping performance decreases greatly with media variation, and was capped at media gradient of the magnitude $0.15$ to keep the running time reasonable. 

The running time of curved ray traversal scales sub-linearly with the number of tetrahedral cells in the mesh, as shown in Figure~\ref{media_scale}(b,c). The traversal performance also scales well with increasing numbers of boundary surfaces (Figure~\ref{media_scale}(d)), which demonstrates the culling efficiency of the tetrahedral mesh. Further discussions about boundary surfaces and performance are given in Section~\ref{others}.   

\subsection{Error analysis of media interpolation}
\label{section_error}
We perform experiments to evaluate the error introduced in our adaptive mesh construction, which resamples the media profile $G$ using a smaller set of points $S$. Assume that the media profile in refractive index $n$ is available as input on a regular grid of points $\mathbf{x}_G$, so that the refractive index $n_G=n(\mathbf{x}_G)$. With Algorithm~\ref{construct} we compute a set of resampled points $S$, their refractive index $n_S$ is computed by trilinear interpolation of $n_G$ on the closest grid points.

After we construct a tetrahedral mesh from the resampled set $S$, the {\em approximated} refractive index $\tilde{n}(\mathbf{x})$ at an arbitrary position $\mathbf{x}$ within the domain is obtained by Barycentric interpolation. The tetrahedral cell that contains $\mathbf{x}$ is located and $\tilde{n}(\mathbf{x})$ is interpolated from $n$ at each vertex of this cell, with Equation \ref{eq:baryncentric}. The approximation error is defined as the difference between the approximated and the original indices of refraction
\begin{equation}
E=n_G-\tilde{n}_G,
\end{equation}
where $\tilde{n}_G=\tilde{n}(\mathbf{x}_G)$. The relative error is
\begin{equation}
E_{rel}= \frac{\|n_G-\tilde{n}_G\|}{\|n_G\|},
\end{equation}
where $\|\cdot\|$ denotes a 2-norm.
The error is a function of the size of $S$, which is controlled by the global
$\sigma$.

Figure~\ref{error_fields} (in the Appendix) shows the approximation error with the profile (A-LU+F). We start from an input grid of $2.09\times10^5$ points, spanning a space of $160$m$\times160$m$\times160$m with $1.25$m grid spacing. By resampling with $\sigma=0.001$, we obtain the resulting $S$ with $23,462$ points. We plotted $S$ in 3D, color-coded by the $n_S$ in Figure~\ref{error_fields}(a), a slice of the original $n_G$ in Figure~\ref{error_fields}(b), the approximated $\tilde{n}_G$ defined by $S$ in Figure~\ref{error_fields}(c), and the error in Figure~\ref{error_fields}(d). With $100$ times fewer points than the input grid, the approximated $\tilde{n}_G$ is able to capture the features of the original $n_G$, and the relative error is below $4\times10^{-4}$. As shown in Figure~\ref{error_fields}(e), the relative error decreases with increasing size of $S$, which is controlled by $\sigma$.

Close approximation of the underlying media profile by the adaptive mesh leads to improved accuracy in the curved ray tracing results. In Figure~\ref{result:hitpoint} (in the Appendix), we quantify the ray tracing accuracy by measuring the spatial locations of ray hit points and the travel distance along the ray trajectories, both of which are crucial for light and sound propagation. Those measurements are compared against a converged piece-wise linear ray stepping result, which is used as the ground truth. With adaptively finer meshes, the approximation errors in ray tracing results decrease along with the approximation errors in the media profiles.

In addition, we perform similar error analysis with two other media profiles (A-HS and A-UW) (see Figure~\ref{result:adaptive_mesh}), which illustrate the capability of our adaptive mesh to capture different profiles with accuracy. The constructed meshes for those profiles are shown in Figure~\ref{result:adaptive_mesh}(e), where finer cells tend to fall in regions of great variations within the media.

\subsection{Comparisons}
\label{results:comparison_cao}
Besides piece-wise linear ray methods, we also compare our algorithm with \cite{Cao2010} on various aspects. By adopting different ray formulation, and by modeling general media with adaptive unstructured mesh, we highlight improvements in performance and accuracy.

We replicate the visual benchmark scenes used in \cite{Cao2010} with the same mesh complexity (see Figure~\ref{render_cao} in the Appendix). Our ray tracer running on a single thread is able to approach the performance of the GPU ray tracer reported in \cite{Cao2010}. This is because of our ray formulation's closed-form intersections with planar surface, which are faster than the costly bisection required by the $n$-linear profile rays used by \cite{Cao2010}. The runtime breakdown in Table~\ref{table:breakdown} shows that bisection takes up a large portion of the traversal time. 

We would also like to point out that our ray formulation has closed-form solution for tangent direction and arc length at any point along the ray curve. These are useful for speeding up boundary interactions as well as absorption and scattering simulation, as explained in Section \ref{section-boundary}. These are not taken into consideration in \cite{Cao2010}.

Moreover, we compare our adaptive mesh with the octree structure used in \cite{Cao2010}. Cao et al.~\cite{Cao2010} built an octree on top of their tetrahedral media to provide the kind of adaptability similar to our tetrahedral mesh formulation. We construct an octree using the the method described in \cite{Cao2010} for the same test profile (A-LU+F) used in Section \ref{section_error} and Figure \ref{error_fields} (in the Appendix). We merge octree nodes according to two thresholds: $\delta$, the threshold of the differences of the indices of refraction of the nodes to be merged, and $\varepsilon$, the threshold of the differences of the gradients. We vary both these thresholds to generate octrees with different number of nodes, and we plot how the relative error changes by reducing the number of nodes (Figure~\ref{error_fields_octree}(e) in the Appendix). Overall the resulting octrees tend to have more nodes when they can achieve the same level of interpolation error as tetrahedral meshes. If comparable number of sample points are used, as the cases plotted in Figure \ref{error_fields} and \ref{error_fields_octree}, the interpolated profile from the octree by finite difference yields visibly less smooth media and larger errors (see Figure~\ref{error_fields_octree}(a-d)).

To estimate the media gradient for the purpose of computing ray curves, we used a regression-based method while Cao et al. \cite{Cao2010} used Green-Gauss method, as discussed in Section \ref{section:gradient}. Although it was mentioned as future work in \cite{Cao2010} that continuity in gradient could potentially remove certain visual artifacts, this improvement is more important for acoustic applications than visual ones. A comparison of the estimated gradient is shown in Figure \ref{result:gradient} (in the Appendix) for acoustic wind profiles A-UW and A-DW.

\subsection{Other considerations}
\label{others}

The design choice of whether to embed the boundary surfaces or not, as discussed in Section~\ref{embed}, depends on whether the resolution of surface tessellations matches the resolution of media variations. Here we show this connection with the geometric representations used for acoustic benchmarks, Christmas and Desert scenes. We tessellate the boundary surfaces in these benchmarks to different resolutions, using the same set of media samples, and construct a different constrained tetrahedral mesh for each resolution. 

As shown in Figure~\ref{result:others}(a) (in the Appendix), there is a particular range of resolution for each scene at which the surface tessellation and the adaptive media mesh resolution match each other; other tessellation levels produce lower-quality mesh with more cells. This effect is even more apparent when we build optimized tetrahedral mesh with a quality threshold measured in the average aspect ratio (Figure~\ref{result:others}(b) in the Appendix).  

On the other hand, when we link boundary surfaces with the media cells they overlap with, rather than embedding them in the mesh, we can see from \ref{result:others}(c) and (d) that a mismatch between surface tessellation and media variation still leads to slower traversal. Even though the mesh is not affected by the surface tessellation in this scenario, the number of surface primitives that overlap each media cell increases with finer surface tessellation, which slows down the traversal.

While constrained mesh construction is more expensive than unconstrained mesh construction, the linking of surfaces also results in significant cost in terms of pre-processing, as shown in Figure \ref{result:others}(e) (in the Appendix). Given a complex media profile with boundaries tessellated at a compatible resolution, the lower traversal time for mesh with embedded boundary surfaces may be worth the extra construction cost.
 
\subsection{Applications on outdoor acoustics}
Ideally, outdoor acoustic simulations model 3D varying media profiles based on temperature and wind profiles, as well as complex natural and man-made boundaries. Existing outdoor acoustic simulation methods either ignore the non-linear media, or simplify the media by reducing the dimension in its variations (e.g. assuming it is simply stratified), or requires long off-line computations. By accelerating the ray models with analytic ray curves and a compact adaptive media mesh, we achieve interactivity with a fully general media profile and complex boundary geometry. Figure~\ref{result:2dray} illustrates the characteristic ray trajectories that we compute for a set of different media conditions. We highlight our method applied to different atmospheric profiles (A-LU,A-LD) and complex outdoor benchmarks Christmas and Desert in Figure~\ref{result:3dray} (in the Appendix).
The resulting ray plots
 display the complex 3D nature of the propagation after multiple interactions with the boundaries (Figure~\ref{result:3dray}) and under wind profiles modified by terrains (Figure~\ref{result:2dray} c,d,g, and h). Our method enables fast generation of those acoustic propagation results.
\section{Limitations and Future work}
There are several limitations to our approach. The first is that the adaptive unstructured mesh is currently precomputed. Therefore, our current implementation is limited to static environments. In dynamic scenes, our approach is limited to the cases where the media property changes do not invalidate the topology of the mesh. Secondly, the efficiency of tracing analytic ray curves depends on the existence of spatial coherence in media. Conceivably there will be a point when a chaotic media has little coherence that the valid range of analytic ray curves reduces to the same with linear ray steps. However, most natural media used for visual and acoustic simulation tends to be fairly coherent and varies smoothly; in these cases tracing analytic ray curves works quite well. 

As future work, we would like to parallelize this approach on a multi-core CPUs or many-core GPUs. Just like linear rays, our analytic ray curves propagates independently from each other, thererfore curved ray traversal is just as amenable to parallelism as linear ray tracing. We would also like to explore modeling of a dynamic media by dynamically adapting the media mesh as in \cite{646238,Cignoni:1994:MMV:197938.197952}, which can be useful for scouting simulation of fluctuating or turbulent media, and of dynamic scenes. Another avenue for future work is to combine our method of simulating refractive propagation with complementary methods that simulate scattering and absorption in participating media. 
\section{Conclusions}
We addressed the challenge of simulating sound and light propagation in large outdoor scenes with general varying media and complex media boundaries. We developed an efficient ray-tracing based algorithm that eliminates the need of making simplifying assumptions about the media variations or the scenes. 

In particular, we traced analytic ray curves that overcome the step size limitation of linear ray approximations, computed closed-form intersections of the ray curves with the scene objects, and constructed adaptive media mesh for efficient representation of the underlying general media profiles. The mesh is also able to conform to the media/objects boundaries, so that surface interactions can be computed seamlessly with media traversal, and the terrain or obstacle-following temperature and wind profiles commonly found in real-world measurements \cite{l1993sound,lamancusa1993ray} can be modeled. 

We highlight the propagation results on outdoor benchmarks with realistic 3D varying atmospheric profiles and complex obstacles, running at near interactive rates on a single CPU core.  Our algorithm enables fast sound simulation in large outdoor scenes and complex environments that were not feasible with previous methods.

\bibliographystyle{IEEEtran}


\section*{Acknowledgment}
This research was supported by ARO Contracts W911NF-10-1-0506, W911NF-12-1-0430, W911NF-13-C-0037, and the National Science Foundation award 1320644.

\bibliography{egbibsample}

\clearpage
\newpage
\pagenumbering{arabic}
\appendices
\section{Derivation of analytic ray curves}
\label{appendix-derivation}
Here we provide the derivation of analytic ray curves based on a locally constant gradient of the propagation speed $c$ and of the squared refractive index $n$. The analytic solutions in various forms have been derived in different context including geometric optics \cite{ob:bornwolf,ob:ghatak,Kravtsov} and computational acoustics \cite{jensen2011,salomons2001}.
\subsection{$c$-linear profile}
\label{section3-1-1}
\noindent When the propagation speed $c$ has a local gradient $\nabla c$, we take the direction of $\nabla c$ as the $z$-axis, and the local media profile can be written as: 
\begin{equation}
\label{eq:c-linear}
c(z)=c_0+\alpha z.
\end{equation}
\noindent From Equation~(\ref{eq:ray-eq-c}) we have
\begin{equation}
\label{eq:c-linear-xyz}
\frac{d}{ds}\left(\frac{1}{c}\frac{dx}{ds}\right)=0,\frac{d}{ds}\left(\frac{1}{c}\frac{dy}{ds}\right)=0,\frac{d}{ds}\left(\frac{1}{c}\frac{dz}{ds}\right)=-\frac{1}{c^2}\frac{\partial c}{\partial z}.
\end{equation}
\noindent We use the following symbols 
\begin{equation}
\label{eq:c-linear-pql}
\xi_0=\frac{1}{c}\frac{dx}{ds},\,\,\,\,\,\,\eta_0=\frac{1}{c}\frac{dy}{ds},\,\,\,\,\,\,\zeta(s)=\frac{1}{c}\frac{dz}{ds},
\end{equation}

\noindent and we can see that $\xi_0$ and $\eta_0$ are constant along the ray trajectory according to Equation~(\ref{eq:c-linear-xyz}). As a result,
\begin{equation}
\label{eq:c-linear-l}
\xi_0^2+\eta_0^2+\zeta^2=\frac{1}{c^2}\left(\left(\frac{dx}{ds}\right)^2+\left(\frac{dy}{ds}\right)^2+\left(\frac{dz}{ds}\right)^2\right)=\frac{1}{c^2}.
\end{equation}

\noindent If we rotate the $x$-$y$ plane around the $z$ axis until $\eta_0$ becomes $0$ and put the origin of the coordinate system at the ray origin, the ray becomes a plane curve lying in the plane formed by the $z$-axis and the initial ray direction at the origin (Figure~\ref{ray_curves} in the paper), which we call the {\it ray plane}. The other axis of the {\it ray plane} is called axis $r$, and that $\xi_0'=\frac{1}{c}\frac{dr}{ds}=\frac{cos\theta_0}{c_0}$, where $\theta_0$ is the angle between initial ray direction and the $r$ axis. 

\noindent In the ray plane, integrating $\frac{dr}{ds}$ along the ray gives
\begin{equation}
\label{eq:c-linear-xy}
r(s_t)=\int_{s_0}^{s_t}\frac{dr}{ds}ds=\xi_0'\int_{s_0}^{s_t} c ds=\xi_0'\int_{z_0}^{z}\frac{dz}{\zeta}.
\end{equation}

\noindent We solve $\zeta$ from Equation~(\ref{eq:c-linear-l}) and plug it into Equation~(\ref{eq:c-linear-xy}), which gives us a circular curve:
\begin{equation}
\label{eq:c-linear-r}
r(z)=\frac{\sqrt{1-\xi_{0}'^{2}c_0^2}-\sqrt{1-\xi_{0}'^{2}\left(c_0+\alpha z\right)^2}}{\xi_0'\alpha}.
\end{equation}

\subsection{$n^2$-linear profile}
\noindent When the squared refractive index $n^2$ has a local gradient $\nabla n^2$, we denote the gradient direction direction as the $z$-axis, so that:
\begin{equation}
\label{eq:n2-linear}
n^2(z)=n_0^2+\alpha z.
\end{equation}
\noindent From Equation~(\ref{eq:ray-eq-n2}), and using a derivation analogous to Equation~(\ref{eq:c-linear-xyz}), (\ref{eq:c-linear-pql}), and (\ref{eq:c-linear-l}), we obtain:

\begin{equation}
\label{eq:n-linear-xyz}
\frac{d}{ds}\left(n\frac{dx}{ds}\right)=0,\frac{d}{ds}\left(n\frac{dy}{ds}\right)=0,\frac{d}{ds}\left(n\frac{dz}{ds}\right)=\frac{\partial n}{\partial z},
\end{equation}
\begin{equation}
\label{eq:n-linear-pql}
\xi_0=n\frac{dx}{ds},\,\,\,\,\,\,\eta_0=n\frac{dy}{ds},\,\,\,\,\,\,\zeta(s)=n\frac{dz}{ds},
\end{equation}
\begin{equation}
\label{eq:n-linear-l}
\xi_0^2+\eta_0^2+\zeta^2=n^2\left(\left(\frac{dx}{ds}\right)^2+\left(\frac{dy}{ds}\right)^2+\left(\frac{dz}{ds}\right)^2\right)=n^2.
\end{equation}
We perform a similar rotation to the ray plane with axis $r$ and $z$, and denote $\xi_0'=n\frac{dr}{ds}=n_0cos\theta_0$. 
As in Equation~(\ref{eq:c-linear-xy}), we obtain:
\begin{equation}
\label{eq:n-linear-xy}
r(s_t)=\int_{s_0}^{s_t}\frac{dr}{ds}ds=\xi_0'\int_{s_0}^{s_t} \frac{ds}{n} =\xi_0'\int_{z_0}^{z}\frac{dz}{\zeta}.
\end{equation}

\noindent We solve $\zeta$ from Equation~(\ref{eq:n-linear-l}) and plug it into Equation~(\ref{eq:n-linear-xy}) to derive the ray trajectory:

\begin{equation}
\label{eq:n2-linear-r}
r(z)=\frac{2\xi_0'}{\alpha}\left(\sqrt{-\xi_{0}'^{2}+n_0^2+\alpha z}-\sqrt{-\xi_{0}'^{2}+n_0^2}\right),
\end{equation}
\noindent which is a parabolic curve.

\section{Gradient estimation solutions}
\label{appendix-gradient}
With linear least square, the estimated gradient from solving Equation (\ref{eq:regression_matrix_weight}) is:
\begin{equation}
\nabla m(\mathbf{x}_0)=\sum\limits_{k=1}^4 \mathbf{p}_k (m(\mathbf{x}_k)-m(\mathbf{x}_0))
\end{equation}

The coefficients, $\mathbf{p}_k$ are:
\begin{equation}
\mathbf{p}_k= \begin{bmatrix} \alpha_{k,1}-\frac{r_{12}}{r_{11}}\alpha_{k,2}+\beta\alpha_{k,3} \\\alpha_{k,2}-\frac{r_{23}}{r_{22}}\alpha_{k,3} \\\alpha_{k,3} \end{bmatrix}
\end{equation}
where 
\begin{subequations}
\begin{equation}
\alpha_{k,1}=\frac{\Delta x_{k}}{r_{11}^2}
\end{equation}
\begin{equation}
\alpha_{k,2}=\frac{1}{r_{22}^2}(\Delta y_{k}-\frac{r_{12}}{r_{11}}\Delta x_{k})
\end{equation}
\begin{equation}
\alpha_{k,3}=\frac{1}{r_{33}^2}(\Delta z_{k}-\frac{r_{23}}{r_{22}}\Delta y_{k}+\beta\Delta x_{k})
\end{equation}
\begin{equation}
\beta=\frac{r_{12}r_{23}-r_{13}r_{22}}{r_{11}r_{22}}
\end{equation}
\end{subequations}
and
\begin{subequations}
\begin{equation}
r_{11}=\sqrt{\sum\limits_{k=1}^4 w_k(\Delta x_{k})^2}
\end{equation}
\begin{equation}
r_{12}=\frac{1}{r_{11}}\sum\limits_{k=1}^4 w_k\Delta x_{k}\Delta y_{k}
\end{equation}
\begin{equation}
r_13=\frac{1}{r_{11}}\sum\limits_{k=1}^4 w_k\Delta x_{k}\Delta z_{k}
\end{equation}
\begin{equation}
r_{22}=\sqrt{\sum\limits_{k=1}^4 w_k(\Delta y_{k})^2-r_{12}^2}
\end{equation}
\begin{equation}
r_{23}=\frac{1}{r_{22}}(\sum\limits_{k=1}^4 w_k\Delta y_{k}\Delta z_{k}-\frac{r_{12}}{r_{11}}\sum\limits_{k=1}^4 w_k \Delta x_{k}\Delta z_{k})
\end{equation}
\begin{equation}
r_{33}=\sqrt{\sum\limits_{k=1}^4 w_k(\Delta z_{k})^2-(r_{13}^2-r_{23}^2)}
\end{equation}
\end{subequations}
where $\Delta (.)=(.)_k-(.)_0$, and $x_{k}$, $y_{k}$, $z_{k}$ are the Cartesian coordinates of $\mathbf{x}_k$.

In contrast, with Green-Gauss gradient estimation as used in \cite{Cao2010}, given a tetrahedral cell with media properties $m$ defined on its vertices $\{m_k,k=1,...,4\}$, the gradient within that cell is given by:
\begin{equation}
\label{eq:baryncentric}
\nabla m=\sum\limits_{k=1}^4\frac{A_k m_k}{T}N_k,
\end{equation}

\noindent where $T$ is the volume of the tetrahedral cell, and $A_k,N_k$ are the area and the normal of the face opposite to vertex $k$, respectively.

This Barycentric interpolation leads to $C^0$-continuity of the media property, $m$, across shared faces, edges, and vertices of neighboring cells. However, there can be discontinuity in the media gradient between neighboring cells.

\begin{figure}
\begin{center}
\subfloat[]{\includegraphics[width=0.45\columnwidth]{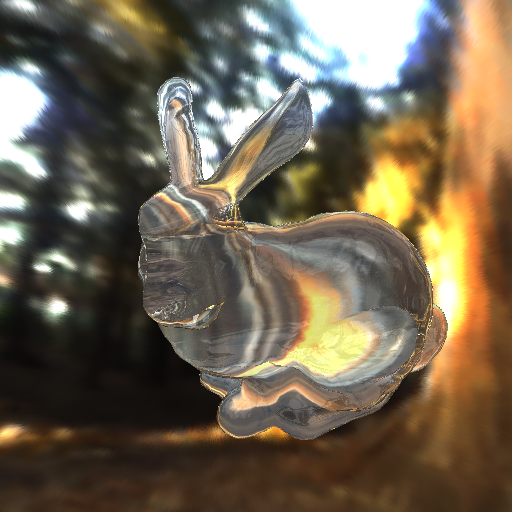}}
\quad
\subfloat[]{\includegraphics[width=0.45\columnwidth]{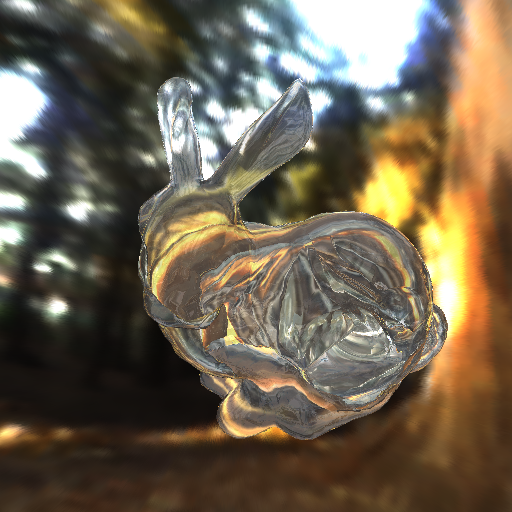}}
\\
\noindent
\subfloat[]{\includegraphics[width=0.45\columnwidth]{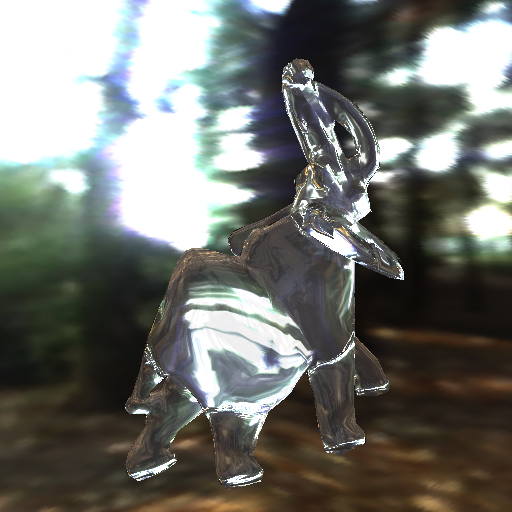}} 
\quad
\subfloat[]{\includegraphics[width=0.45\columnwidth]{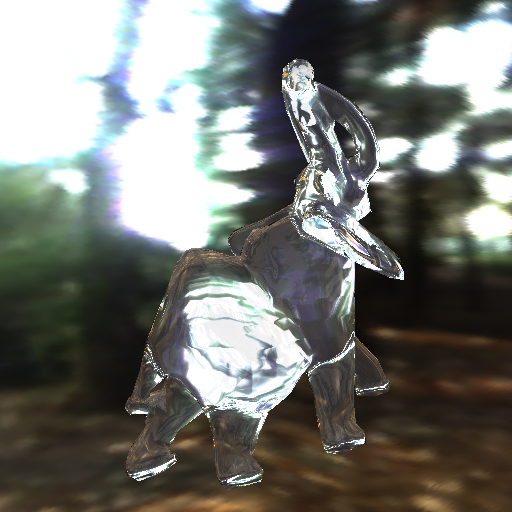}}
\caption{\textbf{Performance comparison} with \cite{Cao2010} on visual benchmarks. We replicate the mesh complexity and scene configuration in Figure 7 of \cite{Cao2010}. \textbf{(a,c)} $n^2$-linear profiles, \textbf{(b,d)} $c$-linear profiles. The medium gradient is along the horizontal direction for Bunny, and along the vertical direction for Elephant. Our curved ray tracer achieved performance of 15 fps for \textbf{(a,b)}, and 8 fps for \textbf{(c,d)}. These frame times achieved with a single CPU thread are within $3\times$ of Cao et al.'s GPU ray tracer \cite{Cao2010}. The key to the efficiency comes from savings of the bisection cost, which can take up to $50\%$ with the $n$-linear ray formulation used in \cite{Cao2010}. See Table~\ref{table:breakdown} for a breakdown of the running time.
}
\label{render_cao}
\end{center}
\end{figure}

\begin{figure*}
\centering
\subfloat[ray curves, 7.3 fps]{\includegraphics[width=0.6\columnwidth]{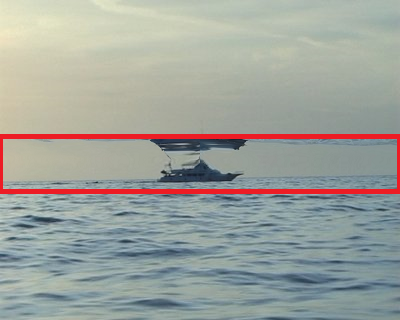}
\label{result:mirage:a}}
\subfloat[ray stepping (size 1.0), 6.92 fps]{\includegraphics[width=0.6\columnwidth]{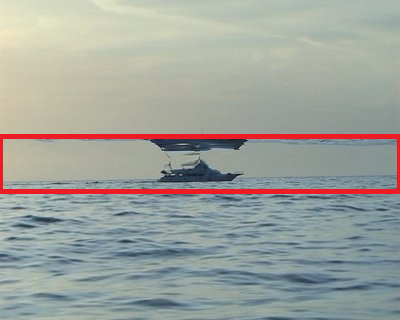}
\label{result:mirage:b}}
\subfloat[ray stepping (size 0.05), 0.22 fps]{\includegraphics[width=0.6\columnwidth]{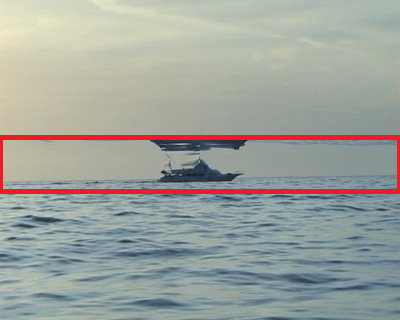}
\label{result:mirage:c}}
\\
\subfloat[photograph of superior mirage]{\includegraphics[width=0.6\columnwidth]{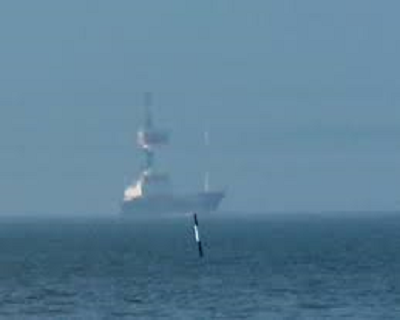}
\label{result:mirage:d}}
\subfloat[diff. between (a) and (b)]{\includegraphics[width=0.6\columnwidth]{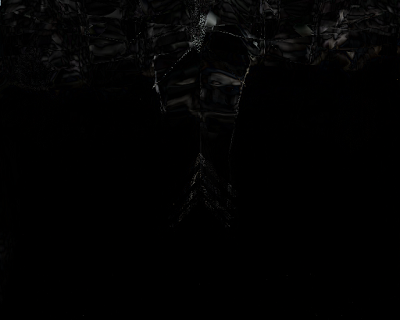}
\label{result:mirage:e}}
\subfloat[diff. between (a) and (c)]{\includegraphics[width=0.6\columnwidth]{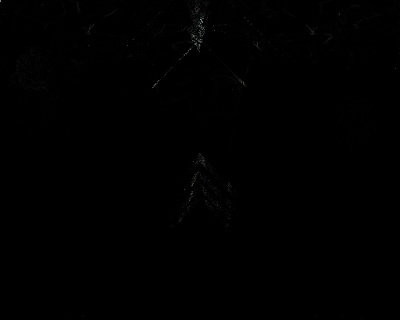}
\label{result:mirage:f}}
\\
\subfloat[ray curves, 8.9 fps]{\includegraphics[width=0.6\columnwidth]{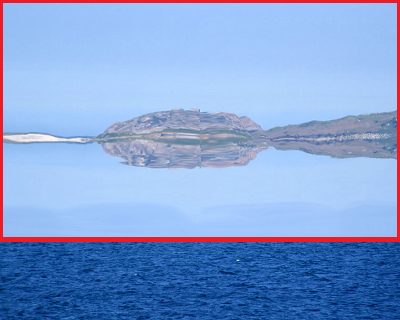}
\label{result:mirage:2-a}}
\subfloat[ray stepping (size 1.0), 8.2 fps]{\includegraphics[width=0.6\columnwidth]{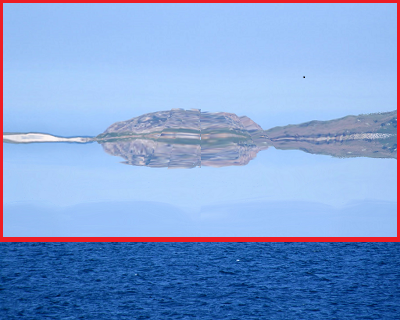}
\label{result:mirage:2-b}}
\subfloat[ray stepping (size 0.05), 0.35 fps]{\includegraphics[width=0.6\columnwidth]{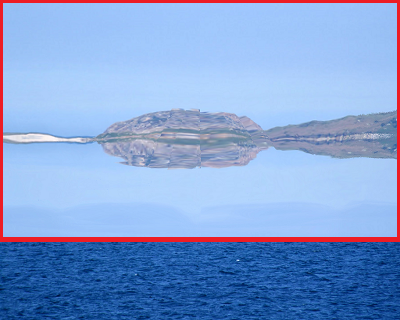}
\label{result:mirage:2-c}}
\\
\subfloat[photograph of inferior mirage]{\includegraphics[width=0.6\columnwidth]{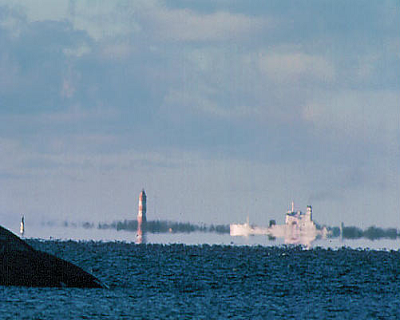}
\label{result:mirage:2-d}}
\subfloat[diff. between (g) and (h)]{\includegraphics[width=0.6\columnwidth]{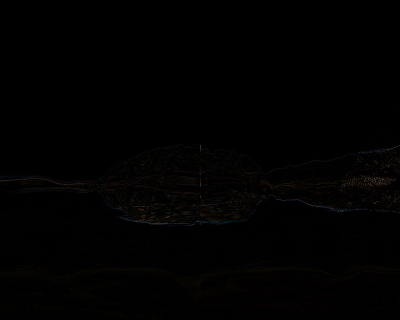}
\label{result:mirage:2-e}}
\subfloat[diff. between (g) and (i)]{\includegraphics[width=0.6\columnwidth]{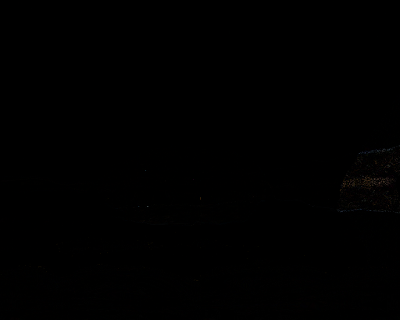}
\label{result:mirage:2-f}}
\caption{\textbf{Same-quality/same-speed comparisons} between curved and linear ray tracing, on visual benchmarks of superior mirages V-SM \textbf{(a-f)} and inferior mirages V-IM \textbf{(g-l)} (see Section \textbf{section6-1} for profile definitions). The atmospheric media is modeled with an adaptive mesh of 28,313 tetrahedral cell, covering a physical volume of $50$m$\times50$m$\times400$m. $512\times512$ rays are traced from the viewer position for each image. \textbf{(a,g)} curved ray tracing results, \textbf{(d,j)} photographs of similar phenomena, \textbf{(b,h)} same-speed comparison, the size of ray steps is chosen to match the performance of curved ray tracing, \textbf{(e,k)} difference images, \textbf{(c,i)} same-quality comparison, the size of ray steps is chosen to match the rendering quality of curved ray tracing, \textbf{(f,l)} difference images. The curved ray tracer is more efficient than ray stepping when rendering at comparable quality. With same speed comparison the artifacts from ray stepping are most visible in areas hit by curved trajectories. All frame rates are measured with single CPU thread.}
\label{result:mirage}
\end{figure*}

\begin{figure*}
\begin{center}
\includegraphics[width=2.0\columnwidth]{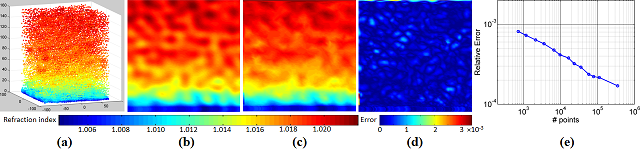}
\caption{\textbf{Approximation error of adaptive meshes.} Approximating the stratified-plus-fluctuation (\textbf{A-LU+F}) atmospheric profile using re-sampled points $S$ containing $100\times$ fewer points than the input profile, and the unstructured mesh that we constructs (Section \ref{section5}). 
\textbf{(a)} The positions of $S$ color-coded by the index of refraction.
\textbf{(b,c)} The original and approximated index of refraction $n_G$, $\tilde{n}_G$ on a slice, respectively.
\textbf{(d)} Absolute error, $|n_G-\tilde{n}_G|$. 
\textbf{(e)} Relative error $ E_{rel}= \|n_G-\tilde{n}_G\|/\|n_G\|$ versus the number of resampled points in $S$. The original grid has $2.09\times10^5$ ($128\times128\times128$)points. 
}
\label{error_fields}
\end{center}
\end{figure*}

\begin{figure*}
\centering
\includegraphics[width=2.0\columnwidth]{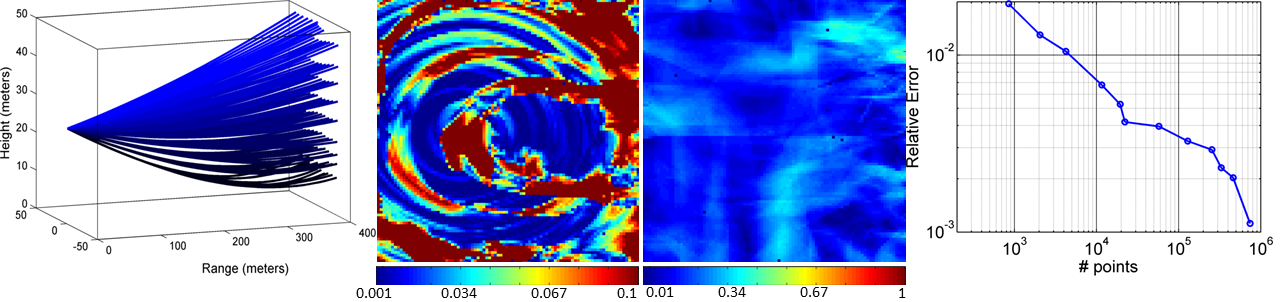}
\caption{\textbf{Approximation error in ray tracing results.} Given the same media profile (\textbf{A-LU+F}) and mesh in Figure \ref{error_fields}, we visualize the resulting errors in ray hit point locations and travel distances along the ray. We use ray stepping with decreasing step size until the ray tracing results converge, and we take the converged results as ground truth.\textbf{(a)} 3D ray curves that we trace, \textbf{(b)} Absolute errors in ray hit point locations, \textbf{(c)} Absolute errors in ray travel distances, \textbf{(d)} the relative error of travel distances decreasing with increasing number of sample points in the adaptive mesh, similar to Figure \ref{error_fields}(e).}
\label{result:hitpoint}
\end{figure*}

\begin{figure*}
\begin{center}
\includegraphics[width=2.0\columnwidth]{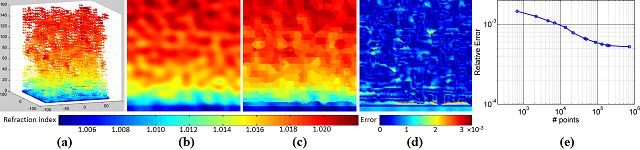}
\caption{\textbf{Compare to approximation error using octree.} We approximate the stratified-plus-fluctuation (\textbf{A-LU+F}) profile using octree, in comparison to the adaptive mesh approximation of our method, analyzed in Figure \ref{error_fields}. We build an octree given the same input media profile on a regular grid of $128\times128\times128$ points, using the same method as \cite{Cao2010}. For the particular octree in (a-d) we use the threshold for differences in indices of refraction $\delta=0.003$ and the threshold for differences in index gradients $\varepsilon=0.0003$, to get similar numbers of samples (26,923) as in the re-sampled points $S$.
\textbf{(a)} The positions of centers of each octree cell, color-coded by the index of refraction.
\textbf{(b,c)} The original and approximated index of refraction $n_G$, $\tilde{n}_G$ on a slice.
\textbf{(d)} Absolute error, $|n_G-\tilde{n}_G|$. 
\textbf{(e)} Relative error $ E_{rel}= \|n_G-\tilde{n}_G\|/\|n_G\|$ versus the number of octree cells. The original grid has $2.09\times10^5$ points.
}
\label{error_fields_octree}
\end{center}
\end{figure*}

\section{Comparison of meshes generated from local gradients of $n,c,n^2$}
For any general media profile, whether given in the propagation speed $c$ or in the refractive index $n$, we could transform the input profile into equivalent profiles of $n$, $c$, or $n^2$ based on the relation $n=c_0/c$. The media gradient in the form of $\nabla n$, $\nabla c$, or $\nabla n^2$ can be computed respectively, and a different adaptive mesh can be constructed using Algorithm \ref{construct} for each of the gradient measures, to be traversed by the $n$-linear, $c$-linear (circular), and $n^2$-linear (parabolic) rays.

\begin{figure*}
\centering
\subfloat[]{\includegraphics[width=2\columnwidth]{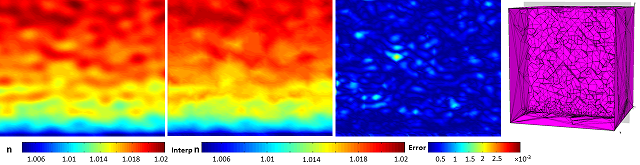}}
\\
\subfloat[]{\includegraphics[width=2\columnwidth]{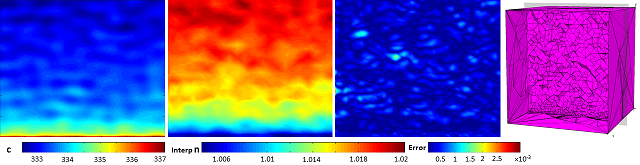}}
\\
\subfloat[]{\includegraphics[width=2\columnwidth]{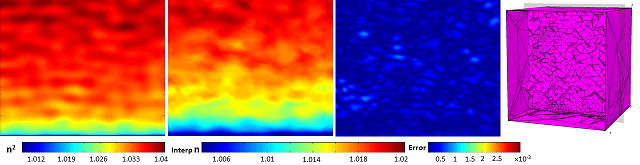}}
\caption{\textbf{Comparisons of 3 analytic ray profiles: upward refractive atmosphere.} With the \textbf{A-LU+F} profile (defined in Section \ref{section6-1}), we compute the same media profile in terms of $c$(sound speed), $n$(acoustic refractive index, with reference $c_0=340 m/s$), and $n^2$, visualized in the leftmost column of \textbf{a,b,c}, respectively. The adaptive meshes constructed according to Algorithm \ref{construct} are shown in the rightmost column of \textbf{a,b,c}, with the control parameters $\sigma=0.001, 0.35, 0.023$, respectively. The control parameters are selected to achieve similar level of approximation error (measured in $n$ and visualized in the second column from right) in the interpolated profiles over the three meshes. The resulting meshes have cell counts of $153867,138965,119670$ respectively, which are roughly on the same level, with the $n^2$-linear profile producing slightly more compact mesh than the other profiles.}
\label{appendix:profiles-upward}
\end{figure*}

\begin{figure*}
\centering
\subfloat[]{\includegraphics[width=2\columnwidth]{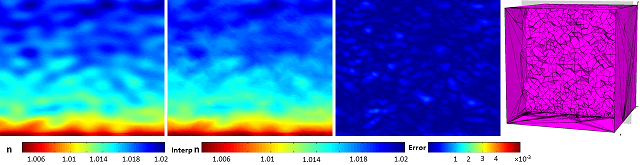}}
\\
\subfloat[]{\includegraphics[width=2\columnwidth]{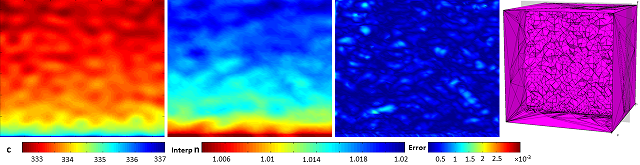}}
\\
\subfloat[]{\includegraphics[width=2\columnwidth]{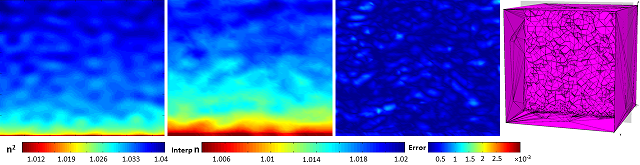}}
\caption{\textbf{Comparisons of 3 analytic ray profiles: downward refractive atmosphere.} With the profile \textbf{A-LD+F} defined in Section \ref{section6-1}, we repeat the experiment in Figure \ref{appendix:profiles-upward}. The equivalent media profiles in terms of $n$, $c$, and $n^2$ are shown in leftmost column in \textbf{a,b,c}, respectively. The adaptive meshes shown in the rightmost column of \textbf{a,b,c} are constructed with the control parameter $\sigma=0.001, 0.3, 0.002$ respectively, achieving similar level of approximation error. The resulting meshes have cell counts of $133735,177958,130759$ respectively. With this downward refracting profile, the mesh sizes are still on the same level, with $n^2$-linear profile producing slightly more compact mesh than the other profiles.}
\label{appendix:profiles-downward}
\end{figure*}

In this Appendix we analyze the approximation errors associated with each of the three kinds of meshes, for the profiles A-LU+F and A-DU+F, in Figure \ref{appendix:profiles-upward} and \ref{appendix:profiles-downward}, respectively. Overall the approximations of the underlying media are at the same accuracy level across different kinds of meshes with comparable size (number of cells). One of the meshes may be better at approximating specific media profiles, but the differences are small. We therefore recommend selecting among the three meshes on a per scene basis, but since the difference is small, $c$-linear and $n^2$-linear profiles may be better choices due to their more efficient boundary intersections.

\begin{figure*}
\centering
\subfloat{\includegraphics[width=0.6\columnwidth]{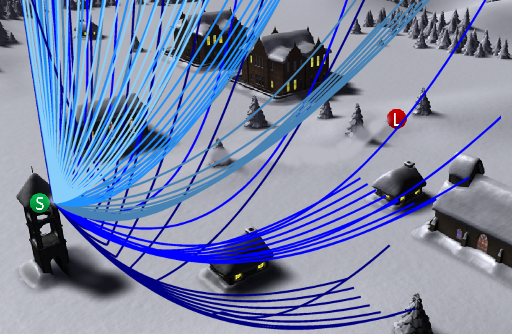}\label{result:3dray:1-a}}
\quad
\subfloat{\includegraphics[width=0.6\columnwidth]{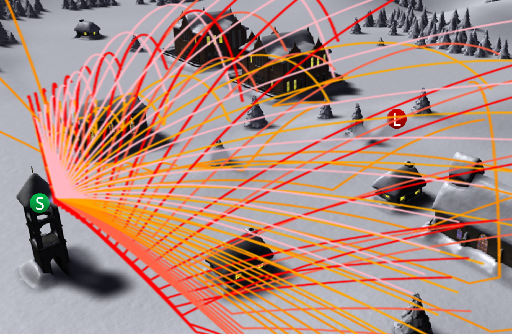}\label{result:3dray:2-a}}
\clearsubcaptcounter
\\
\subfloat{\includegraphics[width=0.6\columnwidth]{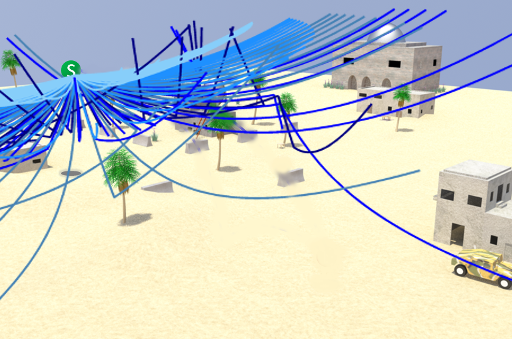}\label{result:3dray:3-a}}
\quad
\subfloat{\includegraphics[width=0.6\columnwidth]{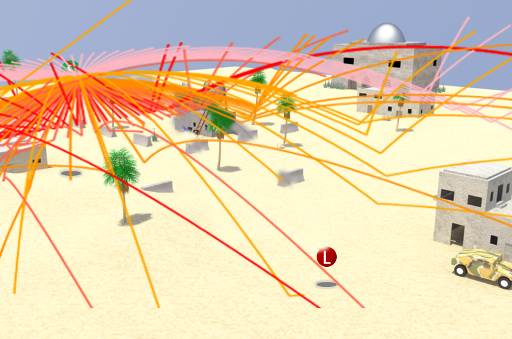}\label{result:3dray:4-a}}
\clearsubcaptcounter
\caption{\textbf{Acoustic propagation.} We compute curved ray trajectories for Christmas and Desert benchmarks. Both upward \textbf{(A-LU+F)} and downward refractive \textbf{(A-LD+F)} atmosphere are simulated. We trace 10K rays for up to 3 surface reflections at 4.5 fps for Desert(m) and 3.8 fps for Christmas(m), respectively. Here we show a representative set of ray paths for each scene and condition. The detailed performance results are listed in Table \ref{table:breakdown}.}
\label{result:3dray}
\end{figure*}

\begin{figure*}
\centering
\subfloat[]{\includegraphics[height=1.5in]{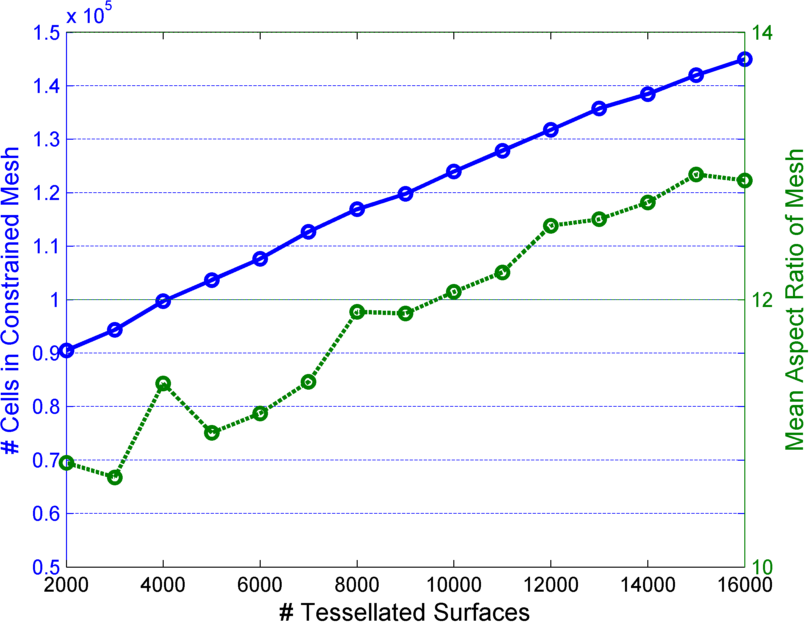}
\label{result:others:a}}
\subfloat[]{\includegraphics[height=1.5in]{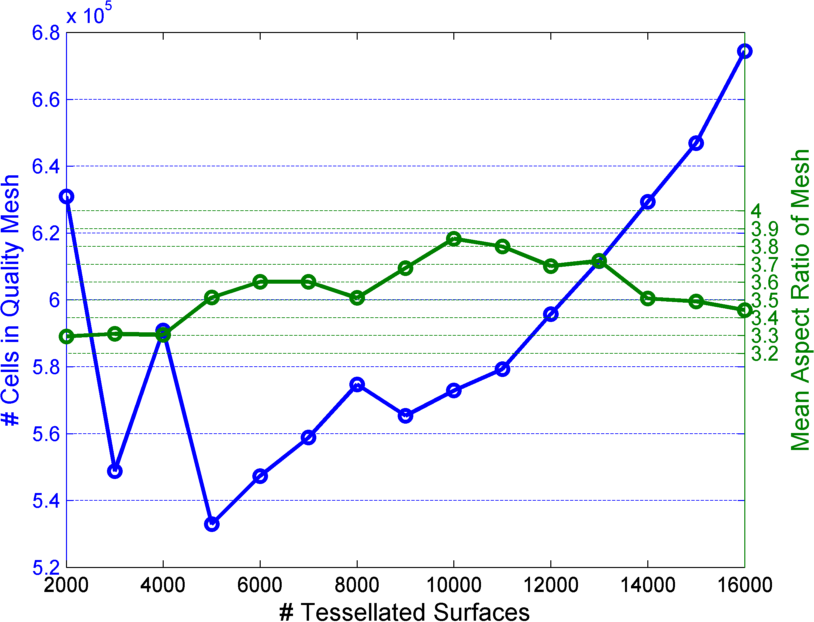}
\label{result:others:b}}
\\
\subfloat[]{\includegraphics[height=1.5in]{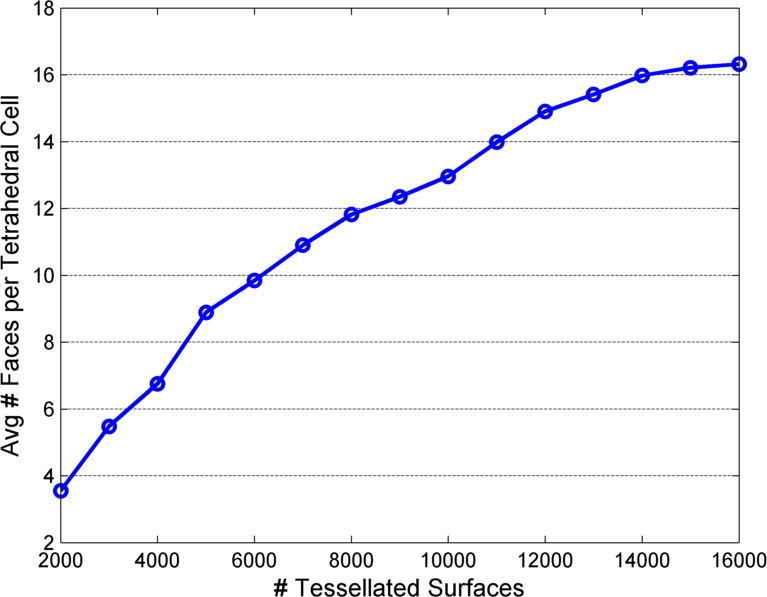}
\label{result:others:c}}
\subfloat[]{\includegraphics[height=1.5in]{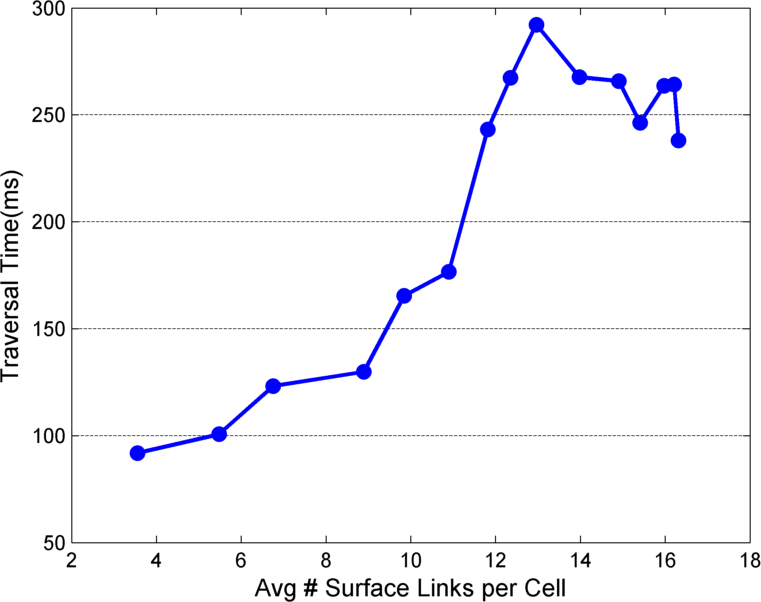}
\label{result:others:d}}
\subfloat[]{\includegraphics[height=1.5in]{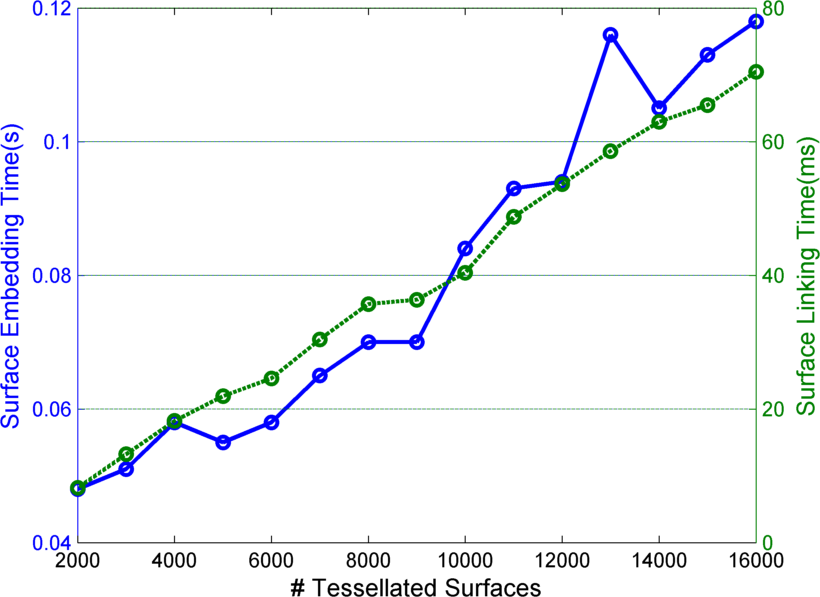}
\label{result:others:e}}
\caption{\textbf{Comparison between embedding and linking boundary surfaces} with regard to the resulting meshes and construction and traversal costs. \textbf{(a)} tessellation of surfaces impacts the sizes and quality of the constrained mesh, the mesh quality reaches a high point (low mean aspect ratio) for surface tessellation that matches the surrounding media sample density. \textbf{(b)} tessellation of surfaces impacts the sizes of quality meshes, which are constrained meshes that are optimized to achieve a quality threshold. With quality constraints, the size of the mesh is most compact when the surface tessellation matches the surrounding media sample density.
\textbf{(c)} tessellation of surfaces impacts the number of surfaces overlapping with each tetrahedral cell, which need to be linked to those mesh cells.  
\textbf{(d)} average number of surface links in turn impacts the traversal performance.
\textbf{(e)} tessellation of surfaces impacts the construction time of both embedding and linking.}
\label{result:others}
\end{figure*}

\begin{figure*}
\centering
\subfloat[upwind over hill, gradient direction, Green-Gauss]{\includegraphics[height=2in]{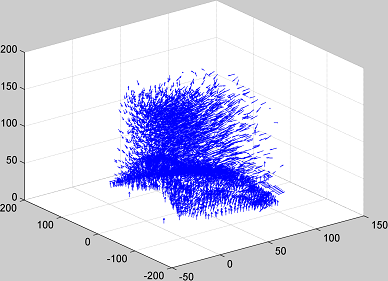}\label{result:gradient:1-a}}
\quad
\subfloat[upwind over hill, gradient direction, regression]{\includegraphics[height=2in]{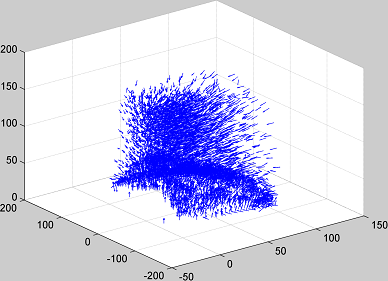}\label{result:gradient:1-b}}
\\
\subfloat[downwind over hill, gradient direction, Green-Gauss]{\includegraphics[height=2in]{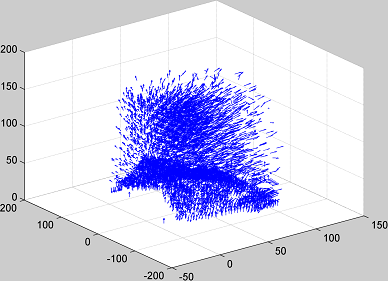}\label{result:gradient:2-a}}
\quad
\subfloat[downwind over hill, gradient direction, regression]{\includegraphics[height=2in]{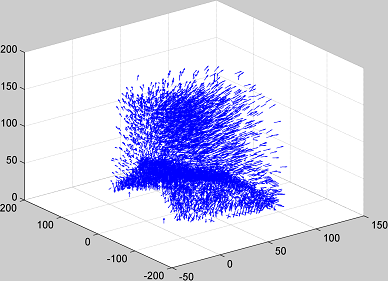}\label{result:gradient:2-b}}
\\
\subfloat[upwind, gradient magnitude, regression]{\includegraphics[width=0.5\columnwidth]{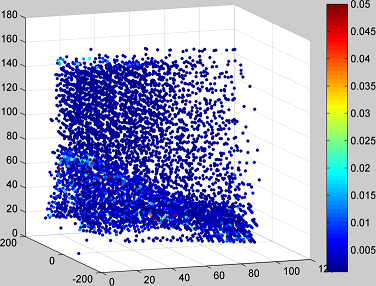}\label{result:gradient:3-a}}
\subfloat[diff. from Green-Gauss]
{\includegraphics[width=0.5\columnwidth]{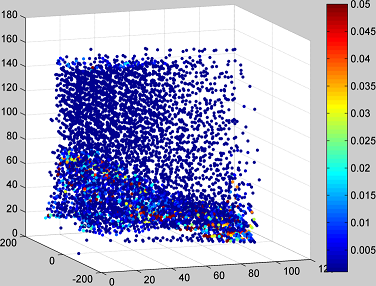}\label{result:gradient:3-b}}
\quad
\subfloat[downwind, gradient magnitude, regression]{\includegraphics[width=0.5\columnwidth]{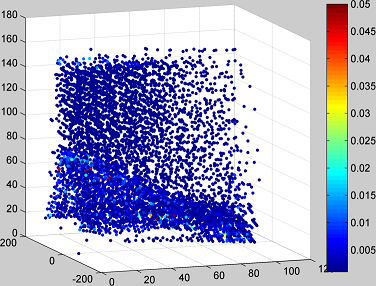}\label{result:gradient:2-d}}
\subfloat[diff. from Green-Gauss]{\includegraphics[width=0.5\columnwidth]{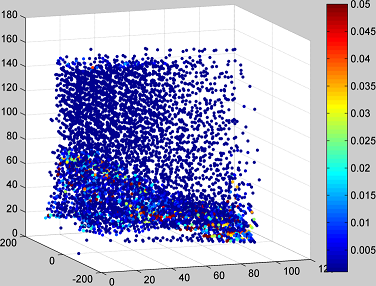}\label{result:gradient:2-d}}
\caption{\textbf{Gradient estimation.} We adopted the regression-based gradient estimation method, which provides better accuracy than Green-Gauss method such as used in \cite{Cao2010}. For acoustic propagation, this additional accuracy is important to avoid artifacts such as false caustics. Here we show side-by-side comparison between the two methods of gradient estimation, applied on the A-UW and A-DW profiles (defined in Section \ref{section6-1} in the paper). The regression method generally produces smoother gradients than Green-Guass in the comparison, computed over the same mesh.}
\label{result:gradient}
\end{figure*}


%

\end{document}